\documentclass[11pt,a4paper,twoside]{article}
\usepackage{graphicx} % Required for inserting images

\usepackage[headheight=25mm,top=30mm,inner=30mm,outer=30mm,marginparwidth=50pt,centering]{geometry}

\usepackage{amsmath,amsfonts,mathrsfs,bbm}
\usepackage{amssymb}
\usepackage{cancel}
\usepackage{slashed}
\usepackage{braket}
\usepackage[normalem]{ulem}

\usepackage{tabularray}
\usepackage{makecell}

\usepackage{comment}
\usepackage{color}
\usepackage[colorlinks=true]{hyperref}
\hypersetup{
    breaklinks=true,
    colorlinks,
    linkcolor={black},
    citecolor={black},
    urlcolor={black}
}

\numberwithin{equation}{section} 

\usepackage{tikz} 
\usetikzlibrary{calc}
\usetikzlibrary{arrows}
\usetikzlibrary{trees}
\usetikzlibrary{decorations.pathmorphing}
\usetikzlibrary{decorations.markings}

\begin{document}

\vspace{36pt}

\begin{center}
{\huge{\bf Groenewold-Moyal twists, integrable spin-chains and AdS/CFT}} 

\vspace{36pt}

Riccardo Borsato and Miguel Garc\'ia Fern\'andez

\vspace{24pt}

{
\small {\it 
Instituto Galego de F\'isica de Altas Enerx\'ias (IGFAE),\\[2pt]
and Departamento de F\'\i sica de Part\'\i culas,\\[2pt]
Universidade de  Santiago de Compostela,\\[2pt]
15705 Santiago de Compostela,  Spain\\[4pt]}
\vspace{12pt}
\texttt{riccardo.borsato@usc.es}, \qquad \texttt{miguelg.fernandez@usc.es}}\\

\vspace{36pt}

{\bf Abstract}
\end{center}
\noindent
We take the first steps to address via integrability the spectral problem of AdS/CFT dual pairs deformed by Groenewold-Moyal twists. In particular, we start by considering a twisted spin-chain that couples, through a Groenewold-Moyal twist deformation, two $\mathfrak{sl}(2)$-invariant spin-chains. We interpret this deformed spin-chain as a deformation of a subsector of the $AdS_3/CFT_2$ spin-chain, but the construction shares qualitative features also with the corresponding deformation of the $AdS_5/CFT_4$ spin-chain, for example. As in similar types of deformations, we show that there exists a certain basis in which the spin-chain Hamiltonian takes a Jordan-block form. At the same time, by working in the basis of eigenstates of the generators used to construct the Groenewold-Moyal twist, the Hamiltonian appears to be diagonalisable and with a deformed spectrum. Employing the method of the Baxter equation, we write down the energy of the ground state and of excited states in a perturbation of the deformation parameter.
We then consider the string-theory side of the duality, where the twist is  realised as a deformation of AdS of the type of Maldacena-Russo-Hashimoto-Itzhaki. We construct a deformation of the usual BMN classical solution, and in the large-$J$ limit we match the leading $\mathcal O(J^{-3})$ term of the energy of the spin-chain ground state with a conserved charge of the classical string solution. Differently from the undeformed setup as well as similar kinds of deformations, we find that the general expression of  this charge of the string sigma-model is non-local, and that it does not correspond to a standard isometry. Nevertheless, it can be computed from the monodromy matrix and it is part of the tower of conserved charges provided by integrability.

\newpage 

%\baselineskip=24pt
%\pagebreak 

\tableofcontents

%%%%%%%%%%%%%%%%%%%%%%%%%%%%%%%%%%%%%%%%%%%%%%%%%%%%%%%%%%%%%%%%%%%%%%%%%%%%%%%%
%%%%%%%%%%%%%%%%%%%%%%%%%%%%%%%%%%%%%%%%%%%%%%%%%%%%%%%%%%%%%%%%%%%%%%%%%%%%%%%%
\section{Introduction}
The Groenewold-Moyal twist is arguably the simplest type of Drinfel'd twist that may be used to construct a non-commutative star-product~\cite{Groenewold:1946kp,Moyal:1949sk}, see e.g.~\cite{Szabo:2001kg} for a review. In the case we are interested in, the star-product of two functions $f(x),g(x)$ is controlled by a deformation parameter $\xi$ and is given by $f(x)\star g(x)=\lim_{y\to x}\exp (\xi (\partial/\partial x^\mu \partial/\partial y^\nu-\partial/\partial x^\nu \partial/\partial y^\mu))f(x)g(y)=f(x)g(x)+\xi(\partial_\mu f\partial_\nu g-\partial_\nu f\partial_\mu g)+\mathcal O(\xi^2)$, where $\mu,\nu$ are two fixed directions identifying a plane in spacetime on which the product fails to be commutative. Via this star-product, one can twist-deform field theories and gauge theories, a subject that attracted a lot of attention in the past and that is now experiencing renewed interest because of new motivations.
The above star-product can be interpreted as an abstract twist of the underlying Hopf algebra, namely as a Drinfel'd twist given by $F_{12}=\exp{(\xi p_\mu\wedge p_\nu)}$ where  $p_\mu$ are spacetime translations, and later we will use this interpretation to twist an integrable spin-chain.

The main interest of this paper is non-commutative deformations of the AdS/CFT correspondence, and their interplay with integrability. Our starting point is the fact that integrable structures appear when studying the spectral problem of maximally supersymmetric dual pairs~\cite{Beisert:2010jr}. For example, in planar $AdS_5/CFT_4$ one identifies integrable spin-chains when calculating anomalous dimensions of gauge-invariant local operators in $\mathcal N=4$ super Yang-Mills~\cite{Minahan:2002ve}, and a classical spectral curve can be shown to encode the spectrum of free strings on the dual $AdS_5\times S^5$ background~\cite{Bena:2003wd}. Via the methods of Thermodynamic Bethe Ansatz and Quantum Spectral Curve~\cite{Gromov:2009tv,Bombardelli:2009ns,Arutyunov:2009ur,Gromov:2013pga}, the spectrum of the theories on the two sides of AdS/CFT can be calculated at finite values of the 't Hooft coupling in the planar limit. Similar integrable structures were identified also in $AdS_4/CFT_3$~\cite{Cavaglia:2014exa} and $AdS_3/CFT_2$~\cite{Chernikov:2025jko,Cavaglia:2025icd}.\footnote{In these cases we are mentioning only the latest references on the Quantum Spectral Curves constructions, because it would be impossible to cite all the works on integrable methods for $AdS_n/CFT_{n-1}$. The case of $AdS_2/CFT_1$ is more complicated from an integrability based approach, but results were obtained also in that case, see e.g.~\cite{Hoare:2014kma,Torrielli:2017nab}.}

Remarkably, several integrable deformations were constructed that are able to deform the sigma-models on $AdS_n\times S^n$ spaces while retaining integrability. In fact, the deformations give rise to type II supergravity backgrounds with deformed geometry and fluxes, that reduce to those of the usual AdS/CFT solutions in the undeformed limit. Integrability is not broken by the deformations because it is possible to construct Lax connections for the deformed sigma-models for general values of the deformation parameter. In particular, the class of ``homogeneous Yang-Baxter deformations''~\cite{Klimcik:2002zj,Klimcik:2008eq,Kawaguchi:2014qwa,Matsumoto:2015jja,vanTongeren:2015soa} is classified and generated by solutions of the classical Yang-Baxter equation on the Lie algebra of isometries of the undeformed background. In the context of $AdS_5/CFT_4$, they were proposed to be dual to non-commutative deformations of $\mathcal N=4$ super Yang-Mills with star-products generated by Drinfel'd twists~\cite{Matsumoto:2014gwa,vanTongeren:2015uha,vanTongeren:2016eeb}.

In this paper we will deal with  the aforementioned Groenewold-Moyal twist, because it yields the prototypical example of a non-commutative deformation of a gauge theory. Our objective will be  to work with this twist in the integrability approach and---lacking an honest derivation of an integrable spin-chain from the twist-deformation of the gauge theory, for example---we will directly twist spin-chains that are relevant for the AdS/CFT correspondence.\footnote{Here in the introduction we will interchangeably use the terms ``deformation'' and ``twist''. In section~\ref{sec:Moyal-spin-chain} we will review the fact that there exist two equivalent pictures, one in which the spin-chain remains periodic and is deformed---in the sense that the Hamiltonian has an explicit dependence on a deformation parameter---and another picture in which the spin-chain is twisted---in the sense that, although the Hamiltonian remains undeformed, now the boundary conditions are not any more periodic  and experience a twist depending on the deformation parameter.} Our starting point will be a spin-chain Hamiltonian that is obtained as the sum of two copies of an $\mathfrak{sl}(2)$-invariant spin-chain, and we will use labels $L$ (left) and $R$ (right) for each of the two copies. This spin-chain can be understood as a subsector of the  spin-chain for strings on $AdS_3\times S^3\times T^4$~\cite{OhlssonSax:2011ms,Borsato:2013qpa}, that we review in section~\ref{sec:AdS3CFT2-spch}. Although it is not possible to interpret this as a subsector of the $\mathcal N=4$ super-Yang-Mills spin-chain for strings on $AdS_5\times S^5$, because in that case there is no closed $\mathfrak{sl}(2)^2$ subsector~\cite{Beisert:2003jj}, this spin-chain and the deformation that we consider still share some qualitative features with the construction that one would have in that case.
In section~\ref{sec:Moyal-spin-chain} we will then build the simplest possible spin-chain with a Groenewold-Moyal twist, in such a way that the twist will couple the $L$ and $R$ copies of the $\mathfrak{sl}(2)^2$-invariant spin-chain. 

The deformation that we consider here is closely related to other Drinfel'd twists that were already studied in the context of AdS/CFT, in particular the dipole deformation of~\cite{Guica:2017mtd} and Jordanian deformations~\cite{Borsato:2022drc,Borsato:2025smn,Driezen:2025dww, Driezen:2025izd}. Let us briefly emphasise the main differences among these constructions. The Groenewold–Moyal and dipole twists are of abelian type, i.e.~these twists are constructed from generators that close into an abelian subalgebra. Strictly speaking, they are not of Drinfel’d–Reshetikhin type~\cite{Reshetikhin:1990ep}, because these commuting generators are not Cartan. In particular, the Groenewold–Moyal twist is generated by two commuting spacetime translations, while the dipole twist combines a spacetime translation with an R-symmetry generator~\cite{Bergman:2000cw}. In contrast, Jordanian twists are qualitatively different. They are based on two non-commuting generators $e,h$ forming a Borel subalgebra $[h,e]=e$, and are therefore non-abelian Drinfel'd twists~\cite{10.1007/BFb0101176,Ogievetsky:1992ph}.

In all these cases the twist  breaks at least some of the Cartan generators that in the undeformed case label the states in the spectral problem, in the sense that the deformed Hamiltonian no longer commutes with them. One is then forced to work with a different basis of the Hilbert space and to label states with eigenvalues of non-Cartan generators that appear in the twists themselves (e.g.~$p_\mu$ in the case of Groenewold-Moyal considered here). In this new basis, the Hamiltonians of the dipole and Jordanian deformed models are diagonalisable and they show  spectra that depend on the deformation parameters entering the twists. The diagonalisation of the Hamiltonians was achieved with the Baxter T--Q approach, under the conjecture that the Baxter equation remains undeformed while the deformation only enters in the expressions for the $Q$-functions. At the same time, one may insist  on working in a basis obtained by taking \emph{finite} linear combinations of eigenstates of the Cartan generators, and would then discover that the Hamiltonian has then the structure of Jordan blocks with undeformed (generalised) eigenvalues. We will confirm this analysis also for the case of the Groenewold-Moyal twist, see sections~\ref{sec:ABA} and~\ref{sec:diagon-Ham}, and we will compute the energy of the ground state and of excited states perturbatively in the deformation parameter.

In the undeformed case, the spin-chain Hamiltonian is typically mapped via AdS/CFT to a conserved charge on the string-theory side, that is $E-J$. Here $E$ is the energy of the string (corresponding to the isometry under translations of global time in $AdS_n$) and $J$ is an angular momentum in $S^n$. In the spin-chain description, $J$ is identified with the length of the chain. It is natural to wonder if it is still possible to associate a conserved charge of the string to the spin-chain Hamiltonian even in the presence of the deformation. As we will argue in section~\ref{sec:string}, the isometries surviving the deformation are not useful to identify such a conserved charge. To address this problem and to start building an AdS/CFT dictionary, we construct a pointlike classical string solution that is a deformation of the BMN one~\cite{Berenstein:2002jq}.  We use this strategy because, already in the undeformed case, see e.g.~\cite{Beisert:2003ea,Kruczenski:2003gt,Stefanski:2004cw}, it was shown that spin-chain  and string-theory calculations for general solutions could be matched in the large-$J$ limit, where fluctuations are suppressed and results can be directly compared. In fact, this strategy was used also in the context of the dipole~\cite{Guica:2017mtd} and Jordanian~\cite{Driezen:2025dww, Driezen:2025izd} deformations to match the energy of the ground state, that in the presence of the deformation takes a non-trivial $J$-dependence. Using the same logic, we will match the leading $\mathcal O(J^{-3})$ term in the large-$J$ expansion of the energy of the ground state of the spin-chain with a conserved charge of the string sigma-model. This match is highly non-trivial, and it allows us to show that the spin-chain Hamiltonian should be dual to a generally \emph{non-local} conserved charge of the string sigma-model, that can be obtained from the monodromy matrix of the classical integrability formulation. To the best of our knowledge, this is the first time that the spin-chain Hamiltonian is mapped via AdS/CFT to a hidden symmetry that is not realised simply as an isometry.

\section{The spin-chain for strings on $AdS_3 \times S^3 \times T^4$}\label{sec:AdS3CFT2-spch}
In this section, we briefly review the spin-chain construction expected to be dual to strings in the massive sector of $AdS_3 \times S^3 \times T^4$. Moreover, we consider all possible subsectors which transform under $\mathfrak{sl}(2)_L \oplus\mathfrak{sl}(2)_R$ in the representation with negative spin $(-1/2,-1/2)$. Finally, we construct the  $XXX_{-1/2}^{\oplus2}$ Hamiltonian and describe the integrability of the model.

\subsection{The $\mathfrak{psu}(1,1|2)_L \oplus \mathfrak{psu}(1,1|2)_R$ spin-chain}
In~\cite{OhlssonSax:2011ms} (see also~\cite{Borsato:2013qpa}), a spin-chain model was constructed to realise the small-coupling regime of the integrable model describing (the AdS/CFT dual of) free strings in the massive sector of $AdS_3 \times S^3 \times T^4$. In particular, it was proposed to be a closed and homogeneous chain with a $\mathfrak{psu}(1,1|2)_L \oplus \mathfrak{psu}(1,1|2)_R$ symmetry algebra.

We start by defining the $\mathfrak{psu}(1,1|2)$ algebra. In what follows, we will consider its complexified version, namely $\mathfrak{psl}(2|2)$. This algebra is defined as the set of all $2|2 \times 2|2$ complex matrices with vanishing supertrace, modulo the center generated by $\mathbb{I}_8$.

The bosonic subalgebra of $\mathfrak{psl}(2|2)$ is $\mathfrak{sl}(2) \oplus\mathfrak{sl}(2)$. We choose the  basis $\{J^3,J^{\pm}\}$ and $\{L^3,L^{\pm}\}$ for each copy of the algebra, respectively, with commutation relations\footnote{In our conventions, the basis for the first $\mathfrak{sl}(2)$ copy of the bosonic algebra differs from the basis  $\{S^3,S^{\pm}\}$ of~\cite{OhlssonSax:2011ms} by the transformation $S^3 \to -J^3,S^+\to J^-, S^- \to - J^+$.}
\begin{align}
    [J^3,J^{\pm}]=\pm J^{\pm} , \quad [J^+,J^-]=-2J^3, \quad
    [L^3,L^{\pm}]=\pm L^{\pm} , \quad [L^+,L^-]=2L^3.
\end{align}
Additionally, there are eight fermionic generators $Q_{a \alpha \dot{\alpha}}$, where each of the three indices can take the two possible values $\pm$. The commutation relations between the bosonic generators and the supercharges are
\begin{align}
    [J^3,Q_{\pm \alpha \dot{\alpha}}]&=\mp \frac{1}{2}Q_{\pm \alpha \dot{\alpha}}, & [J^{\pm},Q_{\pm \alpha \dot{\alpha}}]&=\mp Q_{\mp \alpha \dot{\alpha}},  \\
    [L^3,Q_{a\pm  \dot{\alpha}}]&=\pm \frac{1}{2}Q_{a \pm  \dot{\alpha}}, & [L^{\pm},Q_{a \mp  \dot{\alpha}}]&= Q_{a \pm \dot{\alpha}}.
\end{align}
Finally, the anticommutation relations between the fermionic generators are given by
\begin{align}
    \{Q_{\pm + +} , Q_{\pm - -}\}& =J^{\mp}, & \{Q_{\pm + -},Q_{\pm - +}\}&=-J^{\mp}, \nonumber \\
    \{Q_{+ \pm +},Q_{-\pm -}\}&=\mp L^{\pm} & \{Q_{+\pm -},Q_{- \pm +}\}&=\pm L^{\pm}, \nonumber \\
    \{Q_{+ \pm \pm},Q_{- \mp \mp}\}&=J^3 \pm L^3 & \{Q_{+ \pm \mp},Q_{-\mp \pm}\}&=-J^3 \mp L^3.
\end{align}
The relevant representation for the construction of the spin-chain is the one with weights $(-1/2,1/2)$~\cite{OhlssonSax:2011ms}. The Verma module of this representation is spanned by the states $\ket{\phi_{\alpha}^{(n)}}$ and $\ket{\psi_{\dot{\alpha}}^{(n)}}$, where $\alpha, \dot{\alpha}=\pm$  and $n \in \mathbb{N}$. The action of the bosonic generators is\footnote{In our conventions, the states $\ket{\psi_{\dot{\alpha}}^{(n)}}$ are related to the ones of~\cite{OhlssonSax:2011ms} by the normalisation factor $\frac{1}{\sqrt{n+1}}$.}
\begin{align}
    J^3 \ket{\phi_{\alpha}^{(n)}}& =\left(\frac{1}{2}+n\right)\ket{\phi_{\alpha}^{(n)}}, & J^3 \ket{\psi_{\dot{\alpha}}^{(n)}}& =\left(1+n\right)\ket{\psi_{\dot{\alpha}}^{(n)}}, \nonumber \\
    J^+ \ket{\phi_{\alpha}^{(n)}}& =\left(n+1\right)\ket{\phi_{\alpha}^{(n+1)}}, & J^+ \ket{\psi_{\dot{\alpha}}^{(n)}}& =\left(n+2\right)\ket{\psi_{\dot{\alpha}}^{(n+1)}}, \nonumber \\
    J^- \ket{\phi_{\alpha}^{(n)}}& =n\ket{\phi_{\alpha}^{(n-1)}}, & J^- \ket{\psi_{\dot{\alpha}}^{(n)}}& =n\ket{\psi_{\dot{\alpha}}^{(n-1)}},  
\end{align}    
\begin{align}    
    L^3 \ket{\phi_{\pm}^{(n)}} &= \pm \frac{1}{2} \ket{\phi_{\pm}^{(n)}},& L^+\ket{\phi_{-}^{(n)}} &= \ket{\phi_{+}^{(n)}}, & L^- \ket{\phi_{+}^{(n)}} &= \ket{\phi_{-}^{(n)}}.
\end{align}
Under the first copy of $\mathfrak{sl}(2)$, the states $\ket{\phi_{\alpha}^{(n)}}$ and $\ket{\psi_{\dot{\alpha}}^{(n)}}$ transform in the  $-1/2$ and $-1$ representations, respectively. The index $n$ labels the spin number of the state. Moreover, $\ket{\phi_{\alpha}^{(n)}}$ also transforms in the $1/2$ representation under the second copy of $\mathfrak{sl}(2)$, with the index $\alpha$ labelling the two possible values of the spin quantum number.

In addition, the fermionic generators transform the states $\ket{\phi_{\alpha}^{(n)}}$ into $\ket{\psi_{\dot{\alpha}}^{(n)}}$ and vice versa,
\begin{align}
    Q_{- \pm \dot{\alpha}} \ket{\phi_{\mp}^{(n)}} &= \pm (n+1) \ket{\psi_{\dot{\alpha}}^{(n)}}, &  Q_{+ \pm \dot{\alpha}} \ket{\phi_{\mp}^{(n)}} &= \pm n \ket{\psi_{\dot{\alpha}}^{(n-1)}}, \nonumber \\
    Q_{- \alpha \pm} \ket{\psi_{\mp}^{(n)}} &= \mp \ket{\phi_{\alpha}^{(n+1)}}, & Q_{+ \alpha \pm} \ket{\psi_{\mp}^{(n)}}&=\mp\ket{\phi_{\alpha}^{(n)}}.
\end{align}

The full symmetry algebra of the $AdS_3 \times S^3 \times T^4$ spin-chain is the double copy $\mathfrak{psu}(1,1|2)_L \oplus \mathfrak{psu}(1,1|2)_R$. In particular, one needs to construct two copies of the spin-chain that we will denote by $L$ (left) and $R$ (right). While at all loops $L$ and $R$ excitations do interact~\cite{Borsato:2013qpa}, at weak coupling the scattering between both types of excitations is trivial, and the Hamiltonian reduces to the sum of two $\mathfrak{psu}(1,1|2)$-invariant models with no interaction terms between themselves.

Each copy of the Hamiltonian is a rational $\mathfrak{psu}(1,1|2)$ spin-chain with local Hilbert space given by the Verma module of the $(-1/2,1/2)$ representation. This Hamiltonian coincides with the one-loop dilatation operator of the $\mathfrak{psu}(1,1|2)$ subsector of $\mathcal{N}=4$ super Yang--Mills~\cite{Beisert:2003jj,Beisert:2003yb,Beisert:2007sk}.

Instead of looking for the spectrum of the full spin-chain, one can restrict to a subsector of the Hilbert space in which the Hamiltonian can be diagonalised. In particular, we are interested in the subsectors of states that transform in the $(-1/2,-1/2)$ representation of $\mathfrak{sl}(2)_L \oplus \mathfrak{sl}(2)_R$. The possible sets of states compatible with this symmetry are
\begin{align}
    V_F^{\alpha\beta}=\{\ket{\phi_{\alpha}^{(n)}}_L \otimes \ket{\phi_{\beta}^{(\bar n)}}_R \mid n,\bar n \in \mathbb{N}\}, \label{eq:Hilbert_space} 
\end{align}
with $\alpha$ and $\beta$ fixed. This gives four possibilities, depending on the value of the spin quantum number $\alpha$ and $\beta$. In the following, we will always drop the $\alpha\beta$ indices in $V_F$, always understanding that they are fixed. In the four cases, the residual symmetry algebra is generated by the elements $\{J^3_L,J^\pm_L\}$ and $\{J^3_R,J^\pm_R\}$ of $\mathfrak{sl}(2)_L\oplus \mathfrak{sl}(2)_R\subset\mathfrak{psu}(1,1|2)_L \oplus \mathfrak{psu}(1,1|2)_R$. This $\mathfrak{sl}(2)_L\oplus \mathfrak{sl}(2)_R$ algebra can be understood as the rigid part of the conformal algebra in two dimensions, see~\eqref{eq:map-sl2-conf}.

\subsection{The non-compact $XXX_{-1/2}^{\oplus2}$ spin-chain}
\label{sec:XXX_spin_chain}

The restriction of the full spin-chain to the previous sector leads to a rational $\mathfrak{sl}(2)_L \oplus \mathfrak{sl}(2)_R$ spin-chain, which we denote as $XXX_{-1/2}^{\oplus2}$. Let us now construct the Hamiltonian of this model and present its integrability.

Let $J$ be the length of the chain. The Hilbert space  is given by 
\begin{align}
    \mathcal{H} = V_F^{\otimes J}, \label{eq:full_Hilbert_space}
\end{align}
where $V_F$ is an infinite-dimensional vectorial space defined in~\eqref{eq:Hilbert_space}.

At each site of the chain, the vacuum state is
\begin{align}
    \ket{0}=\ket{\phi_{\alpha}^{(0)}}_L \otimes \ket{\phi_{\beta}^{(0)}}_R.
\end{align}
The excitations are created by acting with the positive-root generators on the vacuum,
\begin{align}
    \ket{(n,\bar n)}:=\ket{\phi_{\alpha}^{(n)}}_L \otimes \ket{\phi_{\beta}^{(\bar n)}}_R =\frac{1}{n! \bar n!}\left(J^+_L\right)^n \left(J^+_R\right)^{\bar n} \ket{0},
\end{align}
where $L$-generators act on the first factor of the tensor product, while the $R$ generators on the second one. In what follows, we work in the oscillator realisation of the $(-1/2,-1/2)$ representation, so that
\begin{align}
    \ket{(n,\bar n)} = (a^\dagger)^n (\bar a^\dagger)^{\bar n} \ket{0},
\end{align}
where $a^\dagger$ and $\bar a^\dagger$ are bosonic creation operators associated with the $L$ and $R$ excitations respectively. We refer the reader to the appendix~\ref{ap:sl_2-sl_2} for a review of the algebra $\mathfrak{sl}(2)_L \oplus \mathfrak{sl}(2)_R$, including the bosonic oscillator realisation.

The $XXX_{-1/2}^{\oplus2}$ spin-chain is defined by the Hamiltonian
\begin{align}
    H = \sum_{p=1}^J \left(h^L_{p,p+1}+h^R_{p,p+1}\right), \quad h_{12}^N=\sum_{j=0}^\infty 2h(j) \mathcal{P}^{N}_{12;j}, \label{eq:undef_ham}
\end{align}
where $h(j)$ is the $j$-th harmonic number and $\mathcal{P}^{N}_{12}$ denotes the projectors defined in~\eqref{eq:projectors} onto the irreducible modules of the decomposition of $V^N_F \otimes V^N_F$~\eqref{eq:moduleVFN}. We will consider a chain with periodic boundary conditions, for which we identify the sites $J+1:=1$.

The Hamiltonian~\eqref{eq:undef_ham} is the sum of two $\mathfrak{sl}(2)$-invariant $XXX_{-1/2}$ models. The Hamiltonian density of the $L$ copy (the same holds for the $R$ copy) acts on a generic two-site state~\cite{Beisert:2003jj} as 
\begin{align}
    h^{L}_{12} \ket{(n_1,\bar n_1);(n_2,\bar n_2)}&=\left( h(n_1) + h(n_2) \right)\ket{(n_1,\bar n_1);(n_2,\bar n_2)} \nonumber \\
    &- \sum_{\substack{k=0 \\ k \neq n_1}}^{n_1+n_2} \frac{1}{|n_1 - k|} \ket{(k,\bar n_1);(n_1+n_2-k,\bar n_2)} \label{eq:harm_act}.
\end{align}
where $\ket{(n_1,\bar n_1);(n_2,\bar n_2)} \in V_F \otimes V_F$ denotes a state with $(n_i,\bar n_i)$ excitations of type $L$ and $R$, respectively, on site $i=1,2$
\begin{align}
    \ket{(n_1,\bar n_1);(n_2,\bar n_2)}= (a_{1}^\dagger)^{n_1}(\bar a_{1}^\dagger)^{\bar n_1}(a_{2}^\dagger)^{n_2}(\bar a_{2}^\dagger)^{\bar n_2} \ket{00}.
\end{align}

The model is invariant under $\mathfrak{sl}(2)_L \oplus \mathfrak{sl}(2)_R$, i.e. the Hamiltonian~\eqref{eq:undef_ham} commutes with $\Delta^{(J)}(x)$ for any $x$ in this algebra. In particular, invariance under $J^3_L$ and $J^3_R$ implies particle number conservation separately for both excitations of type $L$ and $R$,
\begin{align}
    N_L = \sum_{p=1}^J a_p^\dagger a_p, \quad N_R = \sum_{p=1}^J \bar a_p^\dagger \bar a_p.
\end{align}
This property is already manifest in the harmonic action~\eqref{eq:harm_act}, where excitations are only redistributed along the chain, but neither created nor annihilated.

The integrability of the model is based on the existence of an $R$-matrix that satisfies the quantum Yang--Baxter equation (qYBE). Let us consider the following operator
\begin{align}
    R_{12}(u) = \sum_{k,\bar k=0}^\infty R_k(u) R_{\bar k}(u) \mathcal{P}_{12;k,\bar k}, \quad R_{n}(u) = (-1)^{n+1} \frac{\Gamma(-n+u)}{\Gamma(-n-u)} \frac{\Gamma(1-u)}{\Gamma(1+u)}. \label{eq:R-matrix}
\end{align}
From the definition of the projectors $\mathcal{P}_{12;k,\bar k}$~\eqref{eq:projectors}, it is immediate to verify that~\eqref{eq:R-matrix} can be factorised as a product of two $\mathfrak{sl}_2$ invariant rational solutions of the qYBE~\cite{Beisert:2003yb},
\begin{align}
    R_{12}(u) = R_{12}^{L}(u)R_{12}^{R}(u), \quad R_{12}^N(u)=\sum_{j=0}^\infty R_j(u) \mathcal{P}^N_{12;j}
\end{align}
Therefore, since $[R_{12}^{L},R_{12}^{R}]=0$, it follows  that~\eqref{eq:R-matrix} is also a solution of the qYBE. This $R$-matrix is regular, which means that the evaluation of the spectral parameter at $u=0$ yields the permutation operator $P_{12}=P^{L}_{12}P^{R}_{12}$ on $V_F \otimes V_F$,
\begin{align}
     P_{12}\ket{(n_1,\bar n_1);(n_2,\bar n_2)}&=\ket{(n_2,\bar n_2);(n_1,\bar n_1)}, \\ \nonumber
    P^{L}_{12}\ket{(n_1,\bar n_1);(n_2,\bar n_2)}&=\ket{(n_2,\bar n_1);(n_1,\bar n_2)}.
\end{align}
The existence of an $R$-matrix guarantees the construction of an infinite family of commuting charges via the transfer matrix formalism~\cite{Faddeev:1996iy}. In particular, the Hamiltonian~\eqref{eq:undef_ham} can be recovered as
\begin{align}
    h_{12}=P_{12} \frac{d R_{12}(u)}{du}\bigg|_{u=0}.
\end{align}

\subsection{Algebraic Bethe Ansatz for the $XXX_{-1/2}^{\oplus2}$ spin-chain }

The integrability of the model allows for a complete solution of the spin-chain spectral problem. In general, integrable spin-chains with a symmetry algebra of rank greater than two are solved using nested Bethe methods, see~\cite{Belliard:2008di} and references therein. In the present case, however, due to the additive structure of the Hamiltonian~\eqref{eq:undef_ham} and the lack of interaction between the $L$ and $R$ modules, the complexity of finding the spectrum is drastically reduced. In fact, let $\ket{\Psi^N}$ be an eigenstate of the $N=L,R$ copy of~\eqref{eq:undef_ham} with eigenvalue $E_{N}$. Then, the state
\begin{align} 
    \ket{\Psi} = \ket{\Psi^L}\otimes \ket{\Psi^R}
\end{align}
is also an eigenvector of the total Hamiltonian~\eqref{eq:undef_ham} with eigenvalue $E_{L}+E_{R}$. Therefore, finding the spectral problem of the $XXX_{-1/2}^{\oplus2}$ spin-chain reduces to solving the spectrum of $XXX_{-1/2}$. This can be obtained using either the Coordinate or the Algebraic Bethe Ansatz~\cite{Faddeev:1996iy,Faddeev:1994zg,Staudacher_2005} (see also~\cite{Hao:2019cfu,Zhong:2026mfw}).

For completeness, we briefly discuss the application of the Algebraic Bethe Ansatz (ABA) to the $XXX_{-1/2}^{\oplus2}$ spin-chain. The starting point is the vacuum of the theory, given by the $J$-tensor product of the lowest-weight state of $V_F$
\begin{align}
    \ket{\Omega}=\ket{0}^{\otimes J}. \label{eq:vacuum}
\end{align}
Since the model commutes with the number operators $N_L$ and $N_R$, one can construct the excited eigenstates of the theory by fixing the number of $L$ and $R$ excitations above the vacuum. To this end, consider the $\mathfrak{sl}(2)_L \oplus \mathfrak{sl}(2)_R$ invariant $R$-matrix with auxiliary space $\mathbb{C}^4$ and physical space $V_F$ defined in~\eqref{eq:moduleVF}
\begin{align}
    R_{an}= R^L_{an}\otimes R^R_{an}, \quad R^N_{an}=\left(
\begin{array}{cc}
 \frac{2u+1}{2}+(J^3_N)_n & (J^-_N)_n \\
 -(J^+_N)_n & \frac{2u+1}{2}-(J^3_N)_n \\
\end{array}
\right). \label{eq:R-matrix-aux-1/2}
\end{align}
where the index $n$ labels the position where the operator acts. Next, we define the monodromy operator
\begin{align}
    T_a(u) = R_{aJ}(u) \ldots R_{a1}(u) \label{eq:undeformed_trasnfer}
\end{align}
The above operator can be written as a matrix on $\mathbb{C}^2\times \mathbb{C}^2=\mathbb{C}^4$
\begin{align}\label{eq:monod-undef}
    T_a(u)= T^L_{a}(u)\otimes  T^R_{a}(u), \quad T^N_{a}(u)=\left(
\begin{array}{cc}
 A_N(u) & B_N(u) \\
 C_N(u) & D_N(u) \\
\end{array}
\right),
\end{align}
where the entries are operators acting on the physical space. These operators span the so-called Yang-Baxter algebra of the model, with commutation relations defined by the $RTT$-relation
\begin{align}
    R_{a_1, a_2}(u-v) T_{a_1}(u) T_{a_2}(v) = T_{a_2}(v) T_{a_1}(u) R_{a_1, a_2}(u-v). \label{eq:RTT-full}
\end{align}
Moreover, the vacuum~\eqref{eq:vacuum} is an eigenstate of the operators $A_N(u)$ and $D_N(u)$, while it is annihilated by the operator $B_N(u)$,
\begin{align}
    A_N(u) \ket{\Omega} = (u+1)^J \ket{\Omega}, \quad D_N(u) \ket{\Omega} = u^J \ket{\Omega}, \quad B_N(u) \ket{\Omega} =0. \label{eq:acyion-YB-operators-vacuum}
\end{align}
Taking the trace of~\eqref{eq:undeformed_trasnfer} over the auxiliary space gives rise to the transfer matrix
\begin{align}
    \tau(u) = \tau_L(u)\tau_R(u), \quad \tau_N(u) = A_N(u) + D_N(u).  \label{eq:undef_monodromy}
\end{align}
The eigenstates of the transfer matrix are constructed by acting with the $C$ operators on the vacuum~\cite{Faddeev:1996iy}.
In particular, for a given number $(n,\bar n)$ of excitations of type $L$ and $R$, respectively, we have
\begin{align}
    \tau(u) \ket{v_1,\ldots,v_{n};\bar v_1,\ldots,\bar n_{\bar n}} = \Lambda_{n}(u,\{v\})\Lambda_{\bar n}(u,\{\bar v\}) \ket{v_1,\ldots,v_{n};\bar v_1,\ldots,\bar v_{\bar n}}, 
\end{align}
where the eigenvectors and eigenvalues are given by
\begin{align}
    \ket{v_1,\ldots,v_{n};\bar v_1,\ldots,\bar v_{\bar n}} &= C_L(v_1) \ldots C_L(v_{n}) C_R(\bar v_1) \ldots C_R(\bar v_{\bar n}) \ket{\Omega}, \label{eq:undef-eigen}\\
    \Lambda_{n}(u,\{v\})&= (u+1)^J \prod_{k=1}^{n} \frac{u-v_k+1}{u-v_k}+u^J \prod_{k=1}^{n} \frac{u-v_k-1}{u-v_k}, \nonumber \\
    \Lambda_{\bar n}(u,\{\bar v\})&= (u+1)^J \prod_{k=1}^{\bar n} \frac{u-\bar v_k+1}{u-\bar v_k}+u^J \prod_{k=1}^{\bar n} \frac{u-\bar v_k-1}{u-\bar v_k}
    \label{eq:undef-eigenvalue},
\end{align}
provided that both sets of variables $\{v_1,\ldots,v_{n}\}$ and $\{\bar v_1,\ldots,\bar v_{\bar n}\}$, corresponding to the $L$ and $R$ excitations respectively, independently satisfy the Bethe equations
\begin{align}
    \left(\frac{v_k+1}{v_k}\right)^J = \prod_{\substack{j =1 \\ j \neq k}}^{n} \frac{v_k-v_j-1}{v_k-v_j+1}, \quad \left(\frac{\bar v_k+1}{\bar v_k}\right)^J = \prod_{\substack{j =1 \\ j \neq k}}^{\bar n} \frac{\bar v_k-\bar v_j-1}{\bar v_k-\bar v_j+1}.
\end{align}
The eigenstates~\eqref{eq:undef-eigen} are lowest-weight states of the global algebra $\mathfrak{sl}(2)_L \oplus \mathfrak{sl}(2)_R$, since they are annihilated by $\Delta^{(J)}(J^-_L)$ and $\Delta^{(J)}(J^-_R)$. The complete set of eigenvectors of the transfer matrix is obtained by including the descendants of the lowest-weight states
\begin{align}
    \left(\Delta^{(J)}(J^+_L)\right)^p \left(\Delta^{(J)}(J^+_R)\right)^{\bar p} \ket{v_1,\ldots,v_{n};\bar v_1,\ldots,\bar v_{\bar n}}, \quad p,\bar p \in \mathbb{N}. \label{eq:complete_undef_eigen}
\end{align}
The complete set of eigenstates~\eqref{eq:complete_undef_eigen} forms an eigenbasis of the Hamiltonian~\eqref{eq:undef_ham}, with eigenvalues~\cite{Faddeev:1996iy}
\begin{align}
    E_{(n,\bar n)}= -\sum_{k=1}^{n}\frac{1}{v_k^2+v_k}-\sum_{k=1}^{\bar n}\frac{1}{\bar v_k^2+\bar v_k},
\end{align}
where each summation corresponds to the energy of each copy, $L$ and $R$, of the Hamiltonian.

\section{Groenewold-Moyal twist deformation of the $XXX_{-1/2}^{\oplus2}$ spin-chain}
\label{sec:Moyal-spin-chain}
One of the most important algebraic structures underlying the theory of quantum Yang-Baxter integrable models is quasi-triangular Hopf algebras. Therefore, it is natural to construct integrable deformations of quantum models, i.e., deformations that preserve integrability, by deforming quasi-triangular Hopf algebras in a way that preserves their algebraic properties. One paradigmatic example of this class of deformations is Drinfel'd twists~\cite{drinfeld1983constant,Reshetikhin:1990ep,Giaquinto:1994jx} (see also~\cite{Kulish2009}). Here, we consider a particular type of a Drinfel'd twist, that we call the Groenewold-Moyal twist, and employ it to construct an integrable deformation of the $XXX_{-1/2}^{\oplus2}$ spin-chain.

\subsection{The Groenewold-Moyal twist}
Let us consider the following operator acting on $V_F \otimes V_F$,
\begin{align}
    F_{12}=e^{\xi J^-_L \wedge J^-_R}, \label{eq:moyal-twist}
\end{align}
where $J^-_L$ and $J^-_R$ are  $\mathfrak{sl}(2)_L \oplus \mathfrak{sl}(2)_R$ generators. The variable $\xi$ is the deformation parameter, such that in the limit $\xi \to 0$, ~\eqref{eq:moyal-twist} reduces to the identity operator.

We will call the operator~\eqref{eq:moyal-twist}  the Groenewold-Moyal twist because, when composed with the ordinary pointwise product, it gives rise to the Groenewold-Moyal star-product~\cite{Groenewold:1946kp,Moyal:1949sk}, with the non-commutativity involving two spacetime coordinates. It is a Drinfel'd twist in the sense that it satisfies the 2-cocycle condition~\cite{drinfeld1983constant}
\begin{align}
    F_{12} \left(\Delta\otimes\mathbb{I}\right)F =F_{23} \left(\mathbb{I}\otimes\Delta\right)F. \label{eq:2-cocycle}
\end{align}
Moreover, due to the commutativity property $[J^-_L,J^-_R]=0$, the twist is of abelian type. However, it does not belong to the class of Drinfel'd--Reshetikhin twists~\cite{Reshetikhin:1990ep}, since $J^-_L,J^-_R$ are not Cartan. Because of the abelian nature of the twist, it satisfies the factorisation identities 
\begin{align}
    \left(\Delta\otimes\mathbb{I}\right)F &= F_{13} F_{23}, \nonumber \\ 
    \left(\mathbb{I}\otimes\Delta\right)F &= F_{12} F_{13}. \label{eq:F-factori}
\end{align}
These relations allow one to rewrite the cocycle condition~\eqref{eq:2-cocycle} as a constant Yang--Baxter equation
\begin{align}
    F_{12} F_{13} F_{23} = F_{23} F_{13} F_{12}.
\end{align}
Additionally, the quasi-commutativity property of the $R$-matrix, together with~\eqref{eq:F-factori}, implies that the twist satisfies the following intertwining relation
\begin{align}
    R_{12} F_{13} F_{23}=F_{23} F_{13} R_{12}.
\end{align}
The Groenewold-Moyal twist~\eqref{eq:moyal-twist} admits a power expansion in the deformation parameter. The antisymmetrisation of the first-order coefficient, which in this case is already antisymmetric,
\begin{align}
    r_{12} = J^-_L \wedge J^-_R,
\end{align}
is a solution of the classical Yang--Baxter equation. In fact, this property holds for any Drinfel'd twist continuously connected to the identity, and it provides a one-to-one correspondence between Drinfel'd twists and solutions of the classical Yang--Baxter equation~\cite{drinfeld1983constant}, see also~\cite{Giaquinto:1994jx}.

\subsection{The Groenewold-Moyal deformed $XXX_{-1/2}^{\oplus2}$ Hamiltonian}
According to the theory of Drinfel'd twists, the twisted quasi-triangular Hopf algebra possesses a deformed coproduct given by
\begin{align}
    \Delta \rightarrow \Tilde{\Delta}= F_{12} \Delta F_{12}^{-1}, \label{eq:twisted_cop}
\end{align}
together with a corresponding twisted $R$-matrix,
\begin{align}
    R_{12} \rightarrow \Tilde{R}_{12}=F_{21} R_{12} F_{12}^{-1}. \label{eq:twisted_R}
\end{align}
From this new $R$-matrix, one can construct the deformed Hamiltonian by means of the transfer matrix formalism. Let us consider the deformed transfer matrix
\begin{align}
    \Tilde{\tau}(u)=tr_a \left(\Tilde{R}_{aJ}(u)\cdots\Tilde{R}_{a1}(u)\right),
\end{align}
where the subscript $a$ denotes an auxiliary space chosen to be another copy of $V_F$. The above operator is a generating function of mutually commuting operators $\Tilde{Q}_n$, including the Hamiltonian,
\begin{align}
    \Tilde{Q}_{n} = \frac{d^n}{du^n} \ln{\Tilde{\tau}(u)}\bigg|_{u=0}, \quad [\Tilde{Q}_{n},\Tilde{Q}_{m}]=0, \quad \forall n,m \in \mathbb{N}. 
\end{align}
Now, notice that the transformation~\eqref{eq:twisted_R} preserves the regularity property of the $R$-matrix,
\begin{align}
    \Tilde{R}_{12}(0) = R_{12}(0) = P_{12}.
\end{align}
Therefore, the first conserved charge is the shift operator
\begin{align}
    \Tilde{\tau}(0) =e^{\Tilde{Q}_1}= U = P_{12} \cdots P_{J-1,J}.
\end{align}
This property guarantees that the higher conserved charges are boundary-periodic and local operators with an interaction range of $n$. In particular, the deformed Hamiltonian is given by
\begin{align}
    \Tilde{H}:=\Tilde{Q}_2 = \sum_{p=1}^J \Tilde{h}_{p,p+1} \quad \text{such that} \quad \Tilde{h}_{12} = P_{12} \frac{d\Tilde{R}_{12}(u)}{du} \bigg|_{u=0} \quad  \text{and} \quad \Tilde{h}_{J,J+1}= \Tilde{h}_{J,1}\label{eq:deform_H}.
\end{align}
As a result, the deformed Hamiltonian density is related to the undeformed one by a similarity transformation
\begin{align}
    \Tilde{h}_{12} = F_{12} h_{12} F_{12}^{-1}=\Tilde{h}^{L}_{12}+\Tilde{h}^{R}_{12}, \quad \text{with} \quad \Tilde{h}^{N}_{12}= F_{12} h^{N}_{12} F_{12}^{-1}. \label{eq:deform_h_density}
\end{align}
Note that, although we still use labels $L$ and $R$ to indicate which copy of the original Hamiltonian density is deformed, there is now mixing between $L$ and $R$ generators in both summands of the Hamiltonian as a consequence of the twist.

The Hamiltonian density~\eqref{eq:deform_h_density} is invariant under the Groenewold-Moyal-deformed algebra $\mathcal{
U}_{\xi}(\mathfrak{sl}(2)_L \oplus \mathfrak{sl}(2)_R)$. This is a quantum group realising the undeformed $\mathfrak{sl}(2)_L \oplus \mathfrak{sl}(2)_R$ commutation relations with the deformed coproduct~\eqref{eq:twisted_cop}. Explicitly, its action on the basis elements is
\begin{align}
    \tilde{\Delta}(J^{-}_{L}) &= \Delta(J^{-}_{L}), \quad & \tilde{\Delta}(J^{-}_{R}) &= \Delta(J^{-}_{R})\nonumber,
    \\ \tilde{\Delta}(J^3_{L})&= \Delta(J^3_{L}) + \xi J^{-}_L \wedge J^{-}_R, \quad  &\tilde{\Delta}(J^3_{R})&= \Delta(J^3_{R}) + \xi J^{-}_L \wedge J^{-}_R,\nonumber 
\end{align}
\begin{align}
    \tilde{\Delta}(J^{+}_{L}) &= \Delta(J^{+}_{L}) + 2 \xi J^{3}_L \wedge J^{-}_R + \xi^2 \left(J^{-}_L \otimes \left(J^{-}_R\right)^2 +  \left(J^{-}_R\right)^2 \otimes J^{-}_{L}\right),\nonumber \\
    \tilde{\Delta}(J^{+}_{R}) &= \Delta(J^{+}_{R}) + 2 \xi J^{-}_L \wedge J^{3}_R + \xi^2 \left(\left(J^{-}_L\right)^2 \otimes J^{-}_R + J^{-}_R \otimes \left(J^{-}_{L}\right)^2\right). \label{eq:twist_cop_gen}
\end{align}
As it usually happens for Drinfel'd twisted spin-chains, if we consider the total Hamiltonian~\eqref{eq:deform_H}, the boundary term $\tilde{h}_{J1}$ breaks the symmetry under the global quantum Groenewold-Moyal algebra~\eqref{eq:twist_cop_gen}. Following the criteria of~\cite{Borsato:2025smn}, only those elements $\Tilde{\Delta}^{(J)}(x)$ for $x$ in $\mathcal{U}(\mathfrak{sl}(2)_L \oplus \mathfrak{sl}(2)_R)$ that satisfy
\begin{align}
    U^{-1} \Tilde{\Delta}^{(J)}(x) U = \Tilde{\Delta}^{(J)}(x') \quad \text{for some} \quad x' \in \mathcal{U}(\mathfrak{sl}(2)_L \oplus \mathfrak{sl}(2)_R),
\end{align}
will commute with the deformed Hamiltonian, where $U$ is the shift operator. Note that the twisted coproduct on the generators $J^{-}_L$ and $J^{-}_R$ is primitive. Interestingly, the combination $J^3_L-J^3_R$ is also primitive, since
\begin{align}
    \Tilde{\Delta}(J^3_L-J^3_R)= \Delta(J^3_L-J^3_R).
\end{align}
Therefore, these three generators remain symmetries of the deformed model.
\subsection{The Groenewold-Moyal deformed model as an undeformed chain with twisted-boundary conditions}
Every Drinfel'd twist deformation of a spin-chain admits a representation in terms of an undeformed model with twisted boundary conditions~\cite{Kulish2009}. The central object of the construction is the global intertwiner $\Omega$~\cite{Maillet:1996yy} 
\begin{align}
    \Omega= F_{12} F_{(12),3} \cdots F_{(12\ldots J-1),J}, \quad 
    F_{(12\ldots n),n+1}=\left(\Delta^{(n)}\otimes \mathbb{I}\right) F,
\end{align}
which acts as a global twist of the coproduct
\begin{align}
    \tilde{\Delta}^{(J)} = \Omega \Delta^{(J)} \Omega^{-1}.
\end{align}
Moreover, it defines a similarity transformation between the deformed and undeformed Hamiltonian density
\begin{align}
    \tilde{h}_{n,n+1} = \Omega h_{n,n+1} \Omega^{-1}.
\end{align}
The above relation implies that the total deformed Hamiltonian is similar, through $\Omega$, to the following undeformed Hamiltonian with twisted-boundary conditions
\begin{align}
    \mathbb{H} = \sum_{p=1}^{J-1} h_{p,p+1} + S^{-1} h_{J,1} S, \quad \text{with} \quad S=F_{J1}^{-1} \Omega. \label{eq:bt-ham}
\end{align}
The operator $S$, which implements the twisted boundary conditions, is non-local, as it depends on all sites of the spin-chain. In the present case, for the Groenewold-Moyal twist~\eqref{eq:moyal-twist}, a direct computation shows that it takes the form
\begin{align}
     S= e^{2 \xi \left(\left(J^-_L \right)_1S^-_R-\left(J^-_R \right)_1S^-_L\right)},
\end{align}
where we have defined the global negative root generator
\begin{align}
    S^-_N=\Delta^{(J)}(J^-_N), \label{eq:global_root}
\end{align}
and $\left(J^-_N \right)_1$ indicates the action of $J^-_N $ just on the first site of the chain. Additionally, for abelian twists with Hamiltonian~\eqref{eq:bt-ham}, one may construct the following monodromy matrix~\cite{Guica:2017mtd}
\begin{align}
    \mathbb{T}_a(u) = F_a T_a(u) F_a, \quad F_a = F_{a1}^{op} \ldots F_{aJ}^{op}, \label{eq:bt-transfer}
\end{align}
where $T_a(u)$ is the undeformed monodromy matrix, $a$ labels any auxiliary space representation and $F_{an}^{op}$ denotes the composition of the permutation operator with the twist
\begin{align}
    F_{an}^{op} = P_{an} F_{an} P_{an}.
\end{align}
Again, this twisted monodromy matrix is related to the monodromy matrix constructed from the deformed $\tilde{R}_{an}$-matrix via a similarity transformation given by $\Omega$.

\section{Jordan block form of the twisted transfer matrix} \label{sec:ABA}
In this section, we attempt to diagonalise the twisted transfer matrix in the basis of eigenstates of the Cartan generators $J^3_L$ and $J^3_R$. We will show that, in fact, this is not possible and that the Hamiltonian takes the form of Jordan blocks, with (generalised) eigenvalues that remain undeformed.

\subsection{Preliminaries}
We start by constructing the monodromy matrix~\eqref{eq:bt-transfer} with auxiliary space $\mathbb{C}^2\times \mathbb{C}^2$. In the $(1/2,1/2)$ representation of $\mathfrak{sl}(2)_L \oplus \mathfrak{sl}(2)_R$ and in the basis~\eqref{eq:algebra_basis}, the generators read
\begin{align}
    J^3_L = \frac{\sigma^3}{2} \otimes \mathbb{I}_2, \quad  J^+_L=-\sigma^+\otimes \mathbb{I}_2 \quad  J^-_L=\sigma^-\otimes \mathbb{I}_2, \nonumber\\
    J^3_R =  \mathbb{I}_2 \otimes \frac{\sigma^3}{2}, \quad  J^+_R=-\mathbb{I}_2 \otimes\sigma^+  \quad  J^-_R=\mathbb{I}_2 \otimes \sigma^-.
\end{align}
With this, the operator in $F_a$~\eqref{eq:bt-transfer} takes the form
\begin{align}
    F_a = e^{\xi (\mathbb{I}_2 \otimes \sigma^-) \otimes S^-_L}e^{-\xi (\sigma^-\otimes \mathbb{I}_2) \otimes S^-_R},
\end{align}
where $S^-_N$ is the global negative root generator defined in~\eqref{eq:global_root} in the non-compact $(-1/2,-1/2)$ representation. As a matrix on $\mathbb{C}^2 \times \mathbb{C}^2$, the above operator is
\begin{align}
    F_a = \left[\mathbb{I}_2-\xi \begin{pmatrix}
0 & 0 \\
S^-_R & 0
\end{pmatrix}\right] \otimes \left[\mathbb{I}_2+\xi \begin{pmatrix}
0 & 0 \\
S^-_L & 0
\end{pmatrix}\right].
\end{align}
Therefore, the monodromy matrix~\eqref{eq:bt-transfer} is
\begin{align}
    \mathbb{T}_a(u) = \mathbb{T}^L_a(u) \otimes \mathbb{T}^R_a(u), \label{eq:twisted-transfer}
\end{align}
with
\begin{align}
    \mathbb{T}^L_a(u) = \begin{pmatrix}
        A_L-\xi B_L S^-_R & B_L \\
        C_L-\xi (A_L+D_L)S^-_R + \xi^2 B_L (S^-_R)^2 & D_L - \xi B_L S^-_R
    \end{pmatrix},
\end{align}
and
\begin{align}
    \mathbb{T}^R_a(u) = \begin{pmatrix}
        A_R+\xi B_R S^-_L & B_R \\
        C_R+\xi (A_R+D_R)S^-_L + \xi^2 B_R (S^-_L)^2 & D_R + \xi B_R S^-_L
    \end{pmatrix},
\end{align}
where we have omitted the $u$ dependence on the Yang--Baxter operators of the undeformed model and we used~\eqref{eq:monod-undef}.
Importantly, one does not need to worry about the relative position of operators with different labels $L$ and $R$, because they commute.
From the above monodromy matrix, we conclude that the twisted transfer matrix is given by
\begin{align}
    \tilde{\tau}(u)=tr_a(\mathbb{T}_a(u)) = \left(\tau_L(u) -2 \xi B_L(u) S^-_R\right)\left(\tau_R(u) +2 \xi B_R(u) S^-_L\right).  \label{eq:deformed-monodromy}
\end{align}
where $\tau_N(u)$ is the transfer matrix of the $N=L,R$ copy of the undeformed model~\eqref{eq:undef_monodromy}. Notice that before the deformation the transfer matrix factorises into a $L$ and $R$ a component, but now the twisted transfer matrix mixes $L$ and $R$ operators.

One may attempt to diagonalise the transfer matrix~\eqref{eq:deformed-monodromy} making use of an eigenbasis of the Cartan generators of $\mathfrak{sl}(2)_L \otimes \mathfrak{sl}(2)_R$. Note that the vacuum of the undeformed theory is annihilated by the twist~\eqref{eq:moyal-twist}, and it is therefore also the vacuum of the twisted model. In fact, defining the entries of the twisted monodromy matrix~\eqref{eq:twisted-transfer} as new deformed Yang--Baxter operators $\{\tilde{A}_N,\tilde{B}_N,\tilde{C}_N,\tilde{D}_N\}$, one has that~\eqref{eq:vacuum} is a valid reference state of the twisted spin-chain
\begin{align}
    \tilde{B}_N \ket{\Omega} = 0.
\end{align}
Therefore, following the standard route of the Algebraic Bethe Ansatz (ABA), one may attempt to find eigenstates of~\eqref{eq:deformed-monodromy} by acting with $\tilde{C}_N$ operators on $\ket{\Omega}$. An identical approach was employed in the context of Jordanian twist deformations (see~\cite{Borsato:2025smn,Kulish_1997}). However, in those cases the ABA was only able to reproduce the one-magnon sector. The failure to apply the ABA for multiple magnon states was ultimately related to the non-diagonalisability of the models induced by the triangular nature of the Jordanian twist.

As we will see, also in the present case the twisted transfer matrix~\eqref{eq:deformed-monodromy}  defines a non-diagonalisable model when making use of  the eigenbasis of the Cartan generators. Interestingly,  similar results  also arise in the context of other deformations of $\mathcal{N}=4$  super Yang-Mills~\cite{Guica:2017mtd,Ahn:2020zly,Ahn:2022snr,NietoGarcia:2021kgh,NietoGarcia:2022kqi,NietoGarcia:2023jeb}.
We will nevertheless show that it is still possible to construct a sort of ABA, and that this captures the (generalised) eigenvalues of the Hamiltonian.

As a first step towards the construction of a generalised ABA for non-diagonalisable models, we directly compute the (generalised) eigenvectors of the twisted transfer matrix~\eqref{eq:deformed-monodromy} in the eigenbasis of the undeformed model~\eqref{eq:undef-eigen}.
Consider the power expansion of the transfer matrix~\eqref{eq:deformed-monodromy} in the parameter deformation $\xi$,
\begin{align}
    \Tilde{\tau}(u) = \tau(u) + \xi\Tilde{\tau}_1(u) + \xi^2 \Tilde{\tau}_2(u), \label{eq:power-exp-twisted-monodromy}
\end{align}
where $\tau$ is the undeformed transfer matrix~\eqref{eq:undef_monodromy}, while the corrections $\Tilde{\tau}_r$ are given by
\begin{align}
   \Tilde{\tau}_1(u)= 2\tau_L(u) S^-_L B_R(u)-2 B_L(u) \tau_R(u) S^-_R, \quad \Tilde{\tau}_2(u) = -4 B_L(u) S^-_L B_R(u) S^-_R ,\label{eq:deformed_corrections_monodromy}
\end{align}
where we have used that the negative global root generator $S^-_N$ commutes both with the $B_N$ operator and with the undeformed transfer matrix $\tau$. We refer to appendix~\ref{ap:comutation-relations} for the derivation of the commutation relations between the global generators of~$\mathfrak{sl}(2)_L \oplus \mathfrak{sl}(2)_R$ and the Yang-Baxter operators of the undeformed model.

\subsection{On (generalised) eigenstates}
Notice that the action of $\Tilde{\tau}_r$ on a state with $(n,\bar n)$ excitations of type $L$ and $R$, respectively, yields another state with excitation numbers $(n-r,\bar n-r)$. Therefore, the combination $n-\bar n$ is preserved, which explains the symmetry of the deformed model under $J^3_L-J^3_R$. Moreover, the action of~\eqref{eq:deformed_corrections_monodromy} on a state with $n=0$ or $\bar n=0$ vanishes. Thus, eigenstates of the undeformed model with zero excitations of type $L$ or $R$ are also eigenstates of the deformed spin-chain.

In general, because the twist annihilates excitations, the twisted transfer matrix~\eqref{eq:deformed-monodromy} does not commute with $J^3_L$ and $J^3_R$, making it impossible to diagonalise it in a basis with a fixed number of excitations. However, since the twist only annihilates excitations (and it does not create them), one can still attempt to construct eigenstates with a fixed maximum number of $(n,\bar n)$ excitations
\begin{align}
    \ket{\Tilde{\Psi}}_{(n,\bar n)} = \sum_{k=0}^{\min(n,\bar n)} \xi^k \ket{\Tilde{\Phi}}_{(n-k,\bar n-k)} .
\end{align}
Then, we need to solve the eigenvalue equation
\begin{align}
    \Tilde{\tau}(u) \ket{\Tilde{\Psi}}_{(n,\bar n)} = \Tilde{\Lambda}_{n,\bar n}(u) \ket{\Tilde{\Psi}}_{(n,\bar n)},\label{eq:eigen-eq} 
\end{align}
for some eigenvalue $\Tilde{\Lambda}_{n,\bar n}$. Note that this equation must hold order by order in $\xi$. In particular, taking into account the power expansion of the twisted transfer matrix~\eqref{eq:power-exp-twisted-monodromy}, the order $O(\xi^0)$ of~\eqref{eq:eigen-eq} leads to the undeformed eigenvalue equation, 
\begin{align}
    \tau(u) \ket{\Tilde{\Phi}}_{(n,\bar n)} =\Tilde{\Lambda}_{n,\bar n}(u) \ket{\Tilde{\Phi}}_{(n,\bar n)}.
\end{align}
Thus, the state $\ket{\Tilde{\Phi}}_{(n,\bar n)}$ and the eigenvalue $\Tilde{\Lambda}_{n,\bar n}$ coincide with the eigenstate and the eigenvalue of the undeformed model, and under the above assumption that $\ket{\Tilde{\Psi}}_{(n,\bar n)}$ is an eigenstate of the deformed transfer matrix, we can write
\begin{align}
    \ket{\Tilde{\Psi}}_{(n,\bar n)} &= \ket{\Psi}_{(n,\bar n)}+\sum_{k=1}^{\min(n,\bar n)} \xi^k \ket{\Tilde{\Phi}}_{(n-k,\bar n-k)},\nonumber \\  \Tilde{\tau}(u)\ket{\Tilde{\Psi}}_{(n,\bar n)} &= \Lambda_{n}(u) \Lambda_{\bar n}(u) \ket{\Tilde{\Psi}}_{(n,\bar n)}, \label{eq:def-eigenst}
\end{align}
where $\ket{\Psi}_{(n,\bar n)}$ denotes the undeformed eigenstates~\eqref{eq:complete_undef_eigen} and $\Lambda_{n},\Lambda_{\bar n}$ are the undeformed eigenvalues~\eqref{eq:undef-eigenvalue} of the $L$ and $R$ copies, respectively. The corrections $\ket{\Tilde{\Phi}}_{(n-k,\bar n-k)}$ are determined recursively by solving the eigenvalue equation~\eqref{eq:eigen-eq} at order $O(\xi^k)$
\begin{align}
    \tau(u)\ket{\Tilde{\Phi}_{(n-k,\bar n-k)}} &+ \tilde{\tau}_1(u) \ket{\Tilde{\Phi}_{(n-k+1,\bar n-k+1)}} + \tilde{\tau}_2(u) \ket{\Tilde{\Phi}_{(n-k+2,\bar n-k+2)}} = \nonumber\\
    &=\Lambda_{n}(u) \Lambda_{\bar n}(u) \ket{\Tilde{\Phi}_{(n-k,\bar n-k)}}, \quad k=1, \ldots,\min(n,\bar n). \label{eq:correction_eigen-eq}
\end{align}
As already mentioned, undeformed eigenstates~\eqref{eq:complete_undef_eigen} with zero excitations of type $L$ or $R$ are also eigenstates of the twisted model. In addition, all lowest-weight states~\eqref{eq:undef-eigen} are eigenstates of the twisted transfer matrix~\eqref{eq:deformed-monodromy}, since they are annihilated by $S^-_N$. Therefore, all we need to find is the correction to the descendant solutions where both  excitations of type $L$ and $R$ are non-trivial. Interestingly, since $J^-_L$ and $J^-_R$ commute with the deformed Hamiltonian, given any eigenstate  $\ket{\tilde{\Psi}_{(n,\bar n)}}$ with eigenvalue $\Lambda_{n} \Lambda_{\bar n}$, all the states of the form
\begin{align}
    \left(S^-_L\right)^p\left(S^-_R\right)^{\bar p} \ket{\tilde{\Psi}_{(n,\bar n)}}, \quad \text{with} \quad p,\bar p =0,\ldots,\min(n,\bar n).
\end{align}
are also eigenstates of the deformed model with the same eigenvalue $\Lambda_{n} \Lambda_{\bar n}$. 

As we will see, once an  undeformed eigenstate is chosen, the system of equations~\eqref{eq:correction_eigen-eq} does not always admit a solution. This is a signal of the fact that in the basis of eigenstates of the Cartan generators the twisted transfer matrix is not diagonalisable, and then the undeformed eigenstates are associated with generalised eigenstates of the twisted model.

In what follows, we compute the (generalised) eigenvectors of  the twisted transfer matrix $\tilde{\tau}(u)$ for generic values of $u$ in some concrete examples. In particular, we will focus on the subspace spanned by states with at most $n =1$ and $\bar n=1$, as well as $n=2$ and $\bar n=1$ excitations. The strategy to identify the generalised eigenvectors in these cases is the following. Starting from some undeformed eigenstate $\ket{\Psi}$ with excitation numbers $n=2, \bar{n}=1$ or $n=\bar{n}=1$, instead of explicitly solving~\eqref{eq:correction_eigen-eq}, we compute the action of the twisted transfer matrix~\eqref{eq:deformed-monodromy}  on it. Schematically, we have
\begin{align}
    \tilde{\tau}(u)\ket{\Psi}=\Lambda(u) \ket{\Psi} + \ket{\Phi},
\end{align}
where $\ket{\Phi}$ is a state with excitation numbers $n=1, \bar{n}=0$ or $n=\bar{n}=0$. Moreover, one can always write $\ket{\Phi}$ as a linear combination of eigenstates $\{\ket{\phi_i}\}$ of the undeformed model. The key idea is that, in both cases, $\ket{\phi_i}$ has no $R$-excitations, and then, as discussed above, it is also an eigenstate of the twisted transfer matrix. Therefore, one can write 
\begin{align}
    \ket{\Phi} = \sum_{i} \alpha_i(u)\ket{\phi_i}, \quad \text{with} \quad \tilde{\tau}(u) \ket{\phi_i}= \beta_i(u) \ket{\phi_i}.
\end{align}
Now, if all the polynomials $\beta_i(u)$ are different  from $\Lambda(u)$, the state
\begin{align}
    \ket{\tilde{\Psi}} = \ket{\Psi} + \sum_i \frac{\alpha_i(u)}{\Lambda(u)-\beta_i(u)} \ket{\phi_i}
\end{align}
will be a true eigenvector of the twisted transfer matrix with eigenvalue $\Lambda(u)$. Otherwise, if any $\beta_i(u)$ is equal to $\Lambda(u)$, this construction breaks down, and because of the coincidence of the eigenvalues one will have a generalised eigenvector of rank 2 in the Jordan chain of $\ket{\phi_i}$.

Notice that we are interested in finding the (generalised) eigenvectors for generic values of $u$. That is to say, one needs to check whether $\beta_i(u)$ is a polynomial equal or different from $\Lambda(u)$. However, even if the two polynomials are different it may happen that for some value of $u=v_*$, the two polynomials take the same value $\beta_i(v_*) = \Lambda(v_*)$. Then, if $\alpha_i(v_*) \neq 0$, in this limit $\ket{\Psi}$ will be associated with a generalised eigenvector of $\tau(v_*)$. 

Also, suppose there exists one $\beta_j(u)$ equal to $\Lambda(u)$, so that the state
\begin{align}
    \ket{\tilde{\Psi}} = \ket{\Psi} + \sum_{i \neq j} \frac{\alpha_i(u)}{\Lambda(u)-\beta_i(u)} \ket{\phi_i}
\end{align}
is a generalised eigenvector in the Jordan chain of $\ket{\phi_j}$,
\begin{align}
    \tilde{\tau}(u) \ket{\tilde{\Psi}} = \Lambda(u)\ket{\tilde{\Psi}} + \alpha_j(u)\ket{\phi_j}.
\end{align}
Then, if there exists a value $u=v_*$ such that
\begin{align}
    \Lambda(v_*) &\neq \beta_i(v_*) \quad \forall i \neq j, \\
    \alpha_j(v_*) &=0,
\end{align}
the state $\ket{\tilde{\Psi}}$ becomes a true eigenvector of $ \tilde{\tau}(v_*)$.

This discussion implies that, for matrices that depend polynomially on a parameter $u$, the number and size  of their Jordan blocks can change at isolated values of $u$~\cite{NietoGarcia:2023jeb}.
\subsubsection{Example I: $n=\bar n=1$}
The first non-trivial example we consider is the subspace spanned by states with at most $n=\bar n=1$ excitations. In this subspace, $C_L(v_k) C_R(\bar v_{\bar k}) \ket{\Omega}$ gives rise to a lowest-weight eigenstate, and the set of descendant solutions with one  excitation of type $L$ and $R$ is
\begin{align}
    \ket{\Psi_1^{(k)}}=C_L(v_k) S^+_R \ket{\Omega}, \quad
    \ket{\Psi_2^{(\bar k)}}=S^+_L C_R(\bar v_{\bar k}) \ket{\Omega}, \quad
    \ket{\Psi_3}=S^+_L S^+_R \ket{\Omega}, \label{eq:descendant-1L-1R}
\end{align}
where $v_k, \bar v_{\bar k}$ are finite solutions of the Bethe equations in the one-magnon sector
\begin{align}
    \left(\frac{v_k+1}{v_k}\right)^J=1,\quad \left(\frac{\bar v_{\bar k}+1}{\bar v_{\bar k}}\right)^J=1, \quad k, \bar k=1, \ldots, J-1, 
\end{align}
and $S^+_N$ is the global positive root generator of $\mathfrak{sl}(2)_N$
\begin{align}
    S^+_N = \Delta^{(J)}(J^+_N).
\end{align}
The next step is to obtain the (generalised) eigenstates of the twisted transfer matrix~\eqref{eq:deformed-monodromy} associated with the undeformed descendants~\eqref{eq:descendant-1L-1R}.

\subsubsection*{State $\ket{\Psi_1^{(k)}}$:}
We begin by computing the action of $\tilde{\tau}(u)$ on the state $\ket{\Psi_1^{(k)}}$. Note that $\ket{\Psi_1^{(k)}}$ is a lowest-weight state with respect to $\mathfrak{sl}(2)_L$,
\begin{align}
    S^-_L\ket{\Psi_1^{(k)}} = 0.
\end{align}
Therefore, we find
\begin{align}
    \tilde{\tau}(u) \ket{\Psi_1^{(k)}} = \Lambda_{1}(u,v_k)\Lambda_{0}(u) \ket{\Psi_1^{(k)}} -2\xi \Lambda_0(u) B_L(u) S^-_R \ket{\Psi_1^{(k)}}, \label{eq:twisted-monodromy-action_1L-1R}
\end{align}
where $\Lambda_{0}(u)$ and $\Lambda_{1}(u,v_k)$ are the eigenvalues of the undeformed transfer  matrix corresponding to the vacuum and the one-magnon lowest-weight state, respectively
\begin{align}
    \Lambda_{0}(u)=(u+1)^J+u^J, \quad \Lambda_{1}(u,v_k) =(u+1)^J \frac{u-v_k+1}{u-v_k}+u^J \frac{u-v_k-1}{u-v_k}.
\end{align}
Now we simplify the second term on the right-hand side of~\eqref{eq:twisted-monodromy-action_1L-1R}. First, using the $\mathfrak{sl}(2)_N$ commutation relations~\eqref{eq:algebra_basis}, we have
\begin{align}
    S^-_R S^+_R \ket{\Omega} = [S^-_R, S^+_R] \ket{\Omega}= 2 S^3_R \ket{\Omega} = J\ket{\Omega}, \label{eq:S-S+-trick}
\end{align}
which allows one to write
\begin{align}
    B_L(u) S^-_R \ket{\Psi_1^{(k)}} = J B_L(u) C_L(v_k) \ket{\Omega}.
\end{align}
Moreover, the $RTT$ relation~\eqref{eq:RTT-full} for the undeformed model implies the following intertwining relation between the $B_N$ and $C_N$ operators
\begin{align}
    B_N(u) C_N(v) = C_N(v) B_N(u) + \frac{1}{u-v} \left(D_N(v)A_N(u)-D_N(u) A_N(v)\right), \label{eq:intertw-B-C}
\end{align}
from which it follows
\begin{align}
     B_L(u) C_L(v_k) \ket{\Omega} &= C_L(v_k) B_L(u) \ket{\Omega}+\frac{1}{u-v_k} \left(D_L(v_k)A_L(u)-D_L(u) A_L(v_k)\right) \ket{\Omega} = \nonumber \\
     &=\frac{v_k^J}{u-v_k}((u+1)^J-u^J) \ket{\Omega},
\end{align}
where in the last step we have used~\eqref{eq:acyion-YB-operators-vacuum}. With this, the action of the twisted transfer matrix on $\ket{\Psi_1^{(k)}}$ is
\begin{align}
     \tilde{\tau}(u) \ket{\Psi_1^{(k)}} = \Lambda_1(u,v_k) \Lambda_0(u) \ket{\Psi_1^{(k)}} -2\xi J  \Lambda_0(u)\frac{v_k^J}{u-v_k}((u+1)^J-u^J) \ket{\Omega} 
\end{align}
Notice that the vacuum $\ket{\Omega}$ is an eigenvector of $\tilde{\tau}(u)$ with eigenvalue $\Lambda_{0}(u)^2$. Therefore, after some simplifications, one finds that the linear combination
\begin{align}
    \ket{\tilde{\Psi}_1^{(k)}} = \ket{\Psi_1^{(k)}} -2 \xi J v_k^J \ket{\Omega}
\end{align}
is an eigenvector of $\tilde{\tau}(u)$ with eigenvalue $\Lambda_1(u,v_k) \Lambda_0(u)$. 

\subsubsection*{State $\ket{\Psi_2^{(\bar k)}}$:}
We repeat the same calculation with the state $\ket{\Psi_2^{(\bar k)}}$. Now, $\ket{\Psi_2^{(\bar k)}}$ is a lowest-weight state with respect to $\mathfrak{sl}(2)_R$. The action of the twisted transfer matrix is
\begin{align}
    \tilde{\tau}(u) \ket{\Psi_2^{(\bar k)}} = \Lambda_{1}(u,\bar v_k)\Lambda_{0}(u) \ket{\Psi_2^{(\bar k)}} +2\xi \Lambda_0(u) S^-_L B_R(u) \ket{\Psi_2^{(\bar k)}}.
\end{align}
Note that this equation is formally identical to~\eqref{eq:twisted-monodromy-action_1L-1R} under the interchange $L \leftrightarrow R$ and $\xi \to -\xi$. Therefore, using the same identities as before, we end up with
\begin{align}
    \tilde{\tau}(u) \ket{\Psi_2^{(\bar k)}} = \Lambda_1(u,\bar v_{\bar k}) \Lambda_0(u) \ket{\Psi_2^{(\bar k)}}+2\xi J  \Lambda_0(u)\frac{\bar v_{\bar k}^J}{u-\bar v_{\bar k}}((u+1)^J-u^J) \ket{\Omega},
\end{align}
which implies that the deformed state
\begin{align}
    \ket{\tilde{\Psi}_2^{(\bar k)}} = \ket{\Psi_2^{(\bar k)}} +2 \xi J \bar v_{\bar k}^J \ket{\Omega}
\end{align}
is an eigenvector of $\tilde{\tau}(u)$ with eigenvalue $\Lambda_1(u,\bar v_{\bar k}) \Lambda_0(u)$. 
\subsubsection*{State $\ket{\Psi_3}$:}
In this case, the state $\ket{\Psi_3}$ is a descendant with respect to both copies of the algebra. The action of $\tilde{\tau}(u)$ is given by
\begin{align}
    \tilde{\tau}(u) \ket{\Psi_3}&= \Lambda_0(u)^2 \ket{\Psi_3} + 2 \xi \Lambda_0(u) \left(S^-_L B_R(u) - B_L(u) S^-_R\right) \ket{\Psi_3}\nonumber \\
    &- 4 \xi^2 B_L(u) S^-_L B_R(u) S^-_R \ket{\Psi_3}.
\end{align}
The term at order $O(\xi^2)$ vanishes. In fact, using that $S^-_N$ and $B_N(u)$ annihilate the vacuum
\begin{align}
    B_L(u) S^-_L B_R(u) S^-_R \ket{\Psi_3}&= B_L(u) B_R(u) \left[S^-_L,S^+_L\right] \left[S^-_R,S^+_R\right] \ket{\Omega} \nonumber \\
    &\propto B_L(u) B_R(u) \ket{\Omega} = 0.
\end{align}
Moreover, the twisted transfer matrix action at order $O(\xi)$ also vanishes. In particular, using the following commutation relation (see appendix~\ref{ap:comutation-relations})
\begin{align}
    [S^+_N,B_N(u)]=D_N(u)-A_N(u), \label{eq:commutation-S-B}
\end{align}
it follows that
\begin{align}
    \left(S^-_L B_R(u) - B_L(u) S^-_R\right) \ket{\Psi_3} &=\left(\left[S^-_L,S^+_L\right][B_R,S^+_R]-\left[S^-_R,S^+_R\right][B_L,S^+_L]\right) \ket{\Omega} \nonumber \\
    &\propto\left(A_R(u)-A_L(u)+D_L(u)-D_R(u)\right) \ket{\Omega} =0,
\end{align}
where in the last step we have used~\eqref{eq:acyion-YB-operators-vacuum}. With this, we conclude that $ \ket{\Psi_3}$ is an eigenvector of $\tilde{\tau}(u)$ with eigenvalue $\Lambda_0(u)^2$.

In summary, within the subspace spanned by states with at most $n=1$ and $\bar n=1$ excitations, the twisted transfer matrix~\eqref{eq:deformed-monodromy} $\tilde{\tau}(u)$ is fully diagonalisable for any value of $u$. Moreover, the $2J-2$ states associated with the first descendants of either the $L$ or $R$ copy of the algebra receive corrections proportional to the vacuum.

\subsubsection{Example II: $n=2, \bar n=1$}
Now, we consider the subspace spanned by states with up to two excitations of type $L$ and one of type $R$. The set of $(2,1)$-magnon descendant solutions of the untwisted transfer  matrix is given by 
\begin{align}
    \ket{\Psi_1^{(w_1,w_2)}}&=C_L(w_1) C_L(w_2) S^+_R  \ket{\Omega}, &
    \ket{\Psi_2^{(k,\bar k)}}&=C_L(v_k) S^+_L C_R(\bar v_{\bar k}) \ket{\Omega}, \\
    \ket{\Psi_3^{(k)}}&=C_L(v_k) S^+_L S^+_R \ket{\Omega}, &
    \ket{\Psi_4^{(\bar k)}}&=S^+_L S^+_L C_R(\bar v_{\bar k}) \ket{\Omega}, \\
    \ket{\Psi_5}&=S^+_L S^+_L S^+_R \ket{\Omega},
\end{align}
where $w_1,w_2$ are finite solutions of the Bethe equations for the two magnon sector
\begin{align}
    \left(\frac{w_1+1}{w_1}\right)^J= \frac{w_1-w_2-1}{w_1-w_2+1}, \quad \left(\frac{w_2+1}{w_2}\right)^J= \frac{w_2-w_1-1}{w_2-w_1+1},
\end{align}
and $v_k, \bar v_{\bar k}$ for the one-magnon sector
\begin{align}
    \left(\frac{v_k+1}{v_k}\right)^J=1, \quad \left(\frac{\bar v_{\bar k}+1}{\bar v_{\bar k}}\right)^J=1, \quad k, \bar k=1, \ldots, J-1.
\end{align}
\subsubsection*{State $\ket{\Psi_1^{(w_1,w_2)}}$:}
We start with the state $\ket{\Psi_1^{(w_1,w_2)}}$. This is a lowest-weight state with respect to $\mathfrak{sl}(2)_L$. The action of the twisted transfer  matrix is
\begin{align}
    \tilde{\tau}(u)\ket{\Psi_1^{(w_1,w_2)}} = \Lambda_2(u,\{w_i\}) \Lambda_0(u)\ket{\Psi_1^{(w_1,w_2)}} -2 \xi \Lambda_0(u) B_L(u) S^-_R \ket{\Psi_1^{(w_1,w_2)}},
\end{align}
where $\Lambda_2(u,\{w_i\})$ denotes the eigenvalue~\eqref{eq:undef-eigenvalue} associated with the two-magnon lowest-weight state
\begin{align}
    \Lambda_2(u,\{w_1,w_2\}) = (u+1)^J\frac{u-w_1+1}{u-w_1} \frac{u-w_2+1}{u-w_2}+u^J \frac{u-w_1-1}{u-w_1} \frac{u-w_2-1}{u-w_2}.
\end{align}
We now express the contribution at order $O(\xi)$ in terms of the one-magnon eigenvectors of the undeformed transfer matrix and the first descendant of the vacuum 
\begin{align}
    \{S_N^+\ket{\Omega}, C_N(v_k)\ket{\Omega}\}, \quad k=1, \ldots J-1, \label{eq:basis-1-magnon}
\end{align}
with $v_k$ a finite solution of the one-magnon Bethe equations. First, using~\eqref{eq:S-S+-trick}, we obtain
\begin{align}
    B_L(u) S^-_R C_L(w_1) C_L(w_2) S^+_R  \ket{\Omega} = J  B_L(u) C_L(w_1) C_L(w_2) \ket{\Omega}. 
\end{align}
Then, using~\eqref{eq:intertw-B-C} together with the intertwining relations between $A_N$ and $C_N$, and between $D_N$ and $C_N$, derived from~\eqref{eq:RTT-full}, 
\begin{align}
    A_N(u) C_N(v) = \frac{u-v+1}{u-v} C_N(v) A_N(u)-\frac{1}{u-v}C_N(u) A_N(v), \nonumber \\
    D_N(u) C_N(v) = \frac{u-v-1}{u-v} C_N(v) D_N(u)+\frac{1}{u-v}C_N(u) D_N(v), \label{eq:int-AC-DC}
\end{align}
one finds
\begin{align}
    B_L(u) C_L(w_1) C_L(w_2) \ket{\Omega} = \alpha_0 C_L(u) \ket{\Omega} + \alpha_1 C_L(w_1) \ket{\Omega}+\alpha_2 C_L(w_2) \ket{\Omega}, \label{eq:BCC}
\end{align}
where we have defined 
\begin{align}
    \alpha_0 &= -\frac{2 w_1^J w_2^J}{(u-w_1)(u-w_2)}, \\
    \alpha_1 &= \frac{w_2^J (w_2-w_1-1) \left(u^J (u-w_1-1)+(u+1)^J (w_1-u-1)\right)}{(u-w_1) (u-w_2) (w_1-w_2)}, \\ 
    \alpha_2&=\frac{w_1^J (w_1-w_2-1) \left(u^J (w_2-u+1)+(u+1)^J (u-w_2+1)\right)}{(u-w_1) (u-w_2) (w_1-w_2)}.
\end{align}
Finally, we can use the result derived in appendix~\ref{ap:C-action}, which expresses $C_N(u) \ket{\Omega}$ for arbitrary values $u$ in the basis~\eqref{eq:basis-1-magnon},
\begin{align}
    C_N(u) \ket{\Omega} &= \beta_0(u) S^+_N \ket{\Omega} + \sum_{k=1}^{J-1} \beta_k(u) C_N(v_k) \ket{\Omega}, \\
    \beta_0(u)&= \frac{u^J-(u+1)^J}{J}, \quad
    \beta_k(u) = \frac{v_k \left(u^J-(u+1)^J\right)}{J (u-v_k)(1+v_k)^{J-1}}.  \label{eq:action-c}
\end{align}
to rewrite the right-hand side of~\eqref{eq:BCC} as
\begin{align}
     B_L(u) C_L(w_1) C_L(w_2) \ket{\Omega} &= \left(\alpha_0 \beta_0(u)+\alpha_1 \beta_0(w_1)+\alpha_2 \beta_0(w_2)\right) S^+_L \ket{\Omega} + \nonumber \\
     &+\sum_{k=1}^{J-1} \left(\alpha_0 \beta_k(u)+\alpha_1 \beta_k(w_1)+\alpha_2 \beta_k(w_2)\right) C_L(v_k) \ket{\Omega}
\end{align}
Note that the states $S^+_L \ket{\Omega}$ and $C_L(v_k)\ket{\Omega}$ are eigenstates of the twisted transfer  matrix with eigenvalues $\Lambda_0(u)^2$ and $\Lambda_1(u,v_k) \Lambda_0(u)$, respectively. Moreover, as proven in appendix~\ref{ap:proo-dif-eigenv}, all the Bethe states with different magnon numbers or different sets of Bethe roots have different eigenvalues of the transfer matrix. In particular, this implies that
\begin{align}
    \Lambda_2(u,\{w_1,w_2\}) \neq \Lambda_1(u,v_k), \quad \Lambda_2(u,\{w_1,w_2\}) \neq \Lambda_0(u) \label{eq:dif-eigenv}.
\end{align}
Therefore, defining the coefficients
\begin{align}
     \gamma_0 &= \frac{\alpha_0 \beta_0(u)+\alpha_1 \beta_0(w_1)+\alpha_2 \beta_0(w_2)}{\Lambda_2(u,\{w_1,w_2\})-\Lambda_0(u)}=0, \nonumber \\
    \gamma_k &=\frac{\alpha_0 \beta_k(u)+\alpha_1 \beta_k(w_1)+\alpha_2 \beta_k(w_2)}{\Lambda_2(u,\{w_1,w_2\})-\Lambda_1(u,v_k)} = - \frac{2 v_k w_1^J w_2^J}{J(1+v_k)^{J-1}(v_k-w_1)(v_k-w_2)},
\end{align}
one finds that the states
\begin{align}
    \ket{\tilde{\Psi}_1^{(w_1,w_2)}} &= \ket{\Psi_1^{(w_1,w_2)}} - 2 \xi J \gamma_0 S^+_L \ket{\Omega} - 2 \xi J \sum_{k=1}^{J-1} \gamma_k C_L(v_k) \ket{\Omega}, \nonumber \\
    &=\ket{\Psi_1^{(w_1,w_2)}}+ 4 \xi w_1^J w_2^J \sum_{k=1}^{J-1} \frac{v_k}{(1+v_k)^{J-1}(v_k-w_1)(v_k-w_2)} C_L(v_k) \ket{\Omega}.
\end{align}
are eigenstates of $\tilde{\tau}(u)$ with eigenvalue $\Lambda_2(u,\{w_i\}) \Lambda_0(u)$.

\subsubsection*{State $\ket{\Psi_2^{(k,\bar k)}}$:}
The state $\ket{\Psi_2^{(k,\bar k)}}$ is a lowest-weight state with respect to $\mathfrak{sl}(2)_R$. Therefore,
\begin{align}
    \tilde{\tau}(u)\ket{\Psi_2^{(k,\bar k)}} = \Lambda_1(u,v_k)\Lambda_1(u,\bar v_{\bar k}) \ket{\Psi_2^{(k,\bar k)}} + 2 \xi \Lambda_1(u,v_k) S^-_L B_R(u)  \ket{\Psi_2^{(k,\bar k)}}. 
\end{align}
Moreover,
\begin{align}
    S^-_L S^+_L C_L(v_k) \ket{\Omega} = 2 S^3_L C_L(v_k) \ket{\Omega} = (J+2) C_L(v_k) \ket{\Omega}. \label{eq:S3-2-trick}
\end{align}
With this, the term at order $O(\xi)$ simplifies to
\begin{align}
    S^-_L B_R(u)  \ket{\Psi_2^{(k,\bar k)}}&= (J+2) C_L(v_k) B_R(u) C_R(\bar v_{\bar k}) \ket{\Omega} = \nonumber \\
    &=\frac{(J+2)\bar v_{\bar k}^J}{u-\bar v_{\bar k}}\left((u+1)^J-u^J\right)C_L(v_k)\ket{\Omega},
\end{align}
where in the last step we have used~\eqref{eq:intertw-B-C}. Notice that $C_L(v_k)\ket{\Omega}$ is an eigenstate of $\tilde{\tau}(u)$ with eigenvalue $\Lambda_1(u,v_k)\Lambda_0(u)$. Also, from the result derived in appendix~\ref{ap:proo-dif-eigenv}, it is immediate to verify that for finite solutions of the one-magnon Bethe equations all eigenvalues of the transfer  matrix satisfy
\begin{align}
    \Lambda_1(u,v_k)\Lambda_1(u,\bar v_{\bar k}) \neq \Lambda_1(u,v_k)\Lambda_0(u) \quad k,\bar k=1,\ldots,J-1.
\end{align}
Then, computing the quotient
\begin{align}
    \frac{2 \xi \Lambda_1(u,v_k)(J+2)\bar v_{\bar k}^J\left((u+1)^J-u^J\right)}{(u-\bar v_{\bar k})\left(\Lambda_1(u,v_k)\Lambda_1(u,\bar v_{\bar k})-\Lambda_1(u,v_k)\Lambda_0(u)\right)}= 2\xi (J+2) \bar v_{\bar k}^{J},
\end{align}
we conclude that the linear combination
\begin{align}
    \ket{\tilde{\Psi}_2^{(k,\bar k)}} = \ket{\Psi_2^{(k,\bar k)}} + 2\xi (J+2) \bar v_{\bar k}^{J} C_L(v_k) \ket{\Omega}
\end{align}
is an eigenvector of $\tilde{\tau}(u)$ with eigenvalue $\Lambda_1(u,v_k)\Lambda_1(u,\bar v_{\bar k})$.

\subsubsection*{State $\ket{\Psi_3^{(k)}}$:}
In this case, the state $\ket{\Psi_3^{(k)}}$ is a descendant with respect to both copies of the algebra. The action of the twisted transfer matrix~\eqref{eq:deformed-monodromy} is
\begin{align}
    \tilde{\tau}(u) \ket{\Psi_3^{(k)}} &= \Lambda_1(u,v_k)\Lambda_0(u) \ket{\Psi_3^{(k)}} + 2\xi \Lambda_1(u,v_k) S^-_L B_R(u) \ket{\Psi_3^{(k)}} \nonumber \\
    &-2\xi \Lambda_0(u) B_L(u) S^-_R \ket{\Psi_3^{(k)}}-4\xi^2B_L(u) S^-_L B_R(u)S^-_R \ket{\Psi_3^{(k)}}.
\end{align}
The contribution at order $O(\xi^2)$ vanishes 
\begin{align}
  B_L(u) S^-_L B_R(u)S^-_R \ket{\Psi_3^{(k)}} \propto B_R(u) S^-_R S^+_R \ket{\Omega} \propto B_R(u) \ket{\Omega}=0 .
\end{align}
Moreover, using~\eqref{eq:S3-2-trick} and the commutation relation~\eqref{eq:commutation-S-B}, the first contribution at order $O(\xi)$ simplifies to
\begin{align}
    S^-_L B_R(u) \ket{\Psi_3^{(k)}} = (J+2)\left((u+1)^J-u^J\right) C_L(v_k) \ket{\Omega}.  \label{eq:first-order-chi-constribution}
\end{align}
In addition, using the intertwining relations~\eqref{eq:intertw-B-C} and~\eqref{eq:int-AC-DC}, together with the commutation relations (see appendix~\ref{ap:comutation-relations})
\begin{align}
    \left[A_N,S^+_N\right]=-C_N \quad \left[D_N,S^+_N\right]=C_N, \label{eq:commut-A-D-S+}
\end{align}
one can rewrite the remaining term at order $O(\xi)$ as
\begin{align}
    B_L(u) S^-_R \ket{\Psi_3^{(k)}} &= J \left\{\left((u+1)^J-u^J\right)+\frac{(u+1)^J+u^J}{u-v_k}\right\}C_L(v_k) \ket{\Omega} + \nonumber \\
    &+\frac{2 J v_k^J}{v_k-u}C_L(u) \ket{\Omega} + \frac{J v_k^J\left((u+1)^J-u^J\right)}{u-v_k }S^+_L \ket{\Omega}.
\end{align}
Expanding $C_L(u)$ in the undeformed one-magnon eigenbasis~\eqref{eq:action-c} and summing the contribution~\eqref{eq:first-order-chi-constribution}, the action of $\tilde{\tau}(u)$ on $\ket{\Psi_3}$ can be written as
\begin{align}
    \tilde{\tau}(u) \ket{\Psi_3^{(k)}} &= \Lambda_1(u,v_k) \Lambda_0(u)\ket{\Psi_3^{(k)}}+\xi\alpha_0(u) S^+_L\ket{\Omega}+\xi \alpha _k(u) C_L(v_k) \ket{\Omega} \nonumber \\
    &+\xi  \sum_{\substack{r=1 \\r \neq k}}^{J-1} \beta_r(u) C_L(v_r) \ket{\Omega},
\end{align}
where
\begin{align}
    \alpha_0(u) &= \frac{2 (J+2)  \left(u^{2 J}-(u+1)^{2 J}\right) v_k^J}{u-v_k}, \nonumber \\
    \alpha_k(u)&=\frac{4 \left(u^{2 J+1} (-u+2 v_k+1)-2 (J+1) u^J (u+1)^J (u-v_k)+(u+1)^{2 J+1} (u-2 v_k)\right)}{(u-v_k)^2}, \nonumber \\
    \beta_r(u) &= \frac{4 v_r \left(u^{2 J}-(u+1)^{2 J}\right) (v_r+1)^{1-J} v_k^J}{(u-v_r) (u-v_k)}, \quad k \neq r.
\end{align}
Now, note that $C_L(v_k) \ket{\Omega}$ is an eigenstate of $\tilde{\tau}(u)$ with eigenvalue $\Lambda_1(u,v_k) \Lambda_0(u)$. Moreover, all finite one-magnon Bethe roots verify (see appendix~\ref{ap:proo-dif-eigenv})
\begin{align}
   \Lambda_1(u,v_k) =  \Lambda_1(u,v_r) \Leftrightarrow k=r, \quad \text{with} \quad k,r=1,\ldots J-1.
\end{align}
Therefore, computing the coefficients  
\begin{align}
     \frac{ \alpha_0(u)}{\Lambda_1(u,v_k) \Lambda_0(u)-\Lambda_0(u)^2} &= -2(J+2)v_k^J, \nonumber\\
    \frac{\beta_r(u)}{\Lambda_1(u,v_k) \Lambda_0(u)-\Lambda_1(u,v_r) \Lambda_0(u)} &= \frac{4 v_r v_k^J}{(v_r+1)^{J-1}(v_r-v_k)}, \quad r \neq k,
\end{align}
one finds that the state
\begin{align}
    \ket{\tilde{\Psi}_3^{(k)}} =\frac{1}{\alpha_k(u)} \left(\ket{\Psi_3^{(k)}}-2 \xi (J+2) v_k^J S^+_L \ket{\Omega} + 4 \xi v_k^J \sum_{\substack{r=1 \\r \neq k}}^{J-1} \frac{ v_r }{(v_r+1)^{J-1}(v_r-v_k)} C_L(v_r) \ket{\Omega}\right) \label{eq:gen_eigenvec_3}
\end{align}
is a generalised eigenvector of rank 2 with eigenvalue $\Lambda_1(u,v_k)\Lambda_0(u)$ in the Jordan chain of $C_L(v_k) \ket{\Omega}$.

The generalised eigenvector~\eqref{eq:gen_eigenvec_3} diverges at the roots $v_*$ of $\alpha_k(u)$. In this limit, the state $\ket{\tilde{\Psi}_3^{(k)}}$, without the normalisation factor $\alpha_k(u)$, becomes an eigenvector of $\tilde{\tau}(v_*)$.
\subsubsection*{State $\ket{\Psi_4^{(\bar k)}}$:}
Now, the state $\ket{\Psi_4^{(\bar k)}}$ is a lowest-weight state with respect to $\mathfrak{sl}(2)_R$. The action of the twisted transfer matrix is
\begin{align}
    \tilde{\tau}(u) \ket{\Psi_4^{(\bar k)}} = \Lambda_1(u,\bar v_{\bar k}) \Lambda_0(u) \ket{\Psi_4^{(\bar k)}} + 2 \xi \Lambda_0(u) S^-_L B_R(u)  \ket{\Psi_4^{(\bar k)}}.
\end{align}
To compute the contribution at order $O(\xi)$, we first use that
\begin{align}
    S^-_L \left(S^+_L\right)^2 \ket{\Omega} = [ S^-_L, \left(S^+_L\right)^2] \ket{\Omega} = 2 \left(S^+_L S^3_L+S^3_L S^+_L\right) \ket{\Omega} = 2(J+1) S^+_L\ket{\Omega}.
\end{align}
With this,
\begin{align}
    S^-_L  B_R(u) \ket{\Psi_4^{(\bar k)}} &= 2(J+1) S^+_L B_R(u) C_R(\bar v_{\bar k}) \ket{\Omega} = \nonumber \\
    &=2(J+1)\frac{\bar v_{\bar k}^J}{u-\bar v_{\bar k}}((u+1)^J-u^J) S^+_L \ket{\Omega},
\end{align}
where in the last step we have used~\eqref{eq:intertw-B-C}. Therefore, the linear combination
\begin{align}
    \ket{\tilde{\Psi}_4^{(\bar k)}}=\ket{\Psi_4^{(\bar k)}} + 4 \xi (J+1) \bar v_{\bar k}^{J} S^+_L \ket{\Omega}
\end{align}
is an eigenstate of $\tilde{\tau}(u)$ with eigenvalue $\Lambda_1(u,\bar v_{\bar k}) \Lambda_0(u)$.
\subsubsection*{State $\ket{\Psi_5}$:}
Finally, the state $\ket{\Psi_5}$ is a descendant with respect to both copies of the algebra. The action of the twisted transfer matrix is
\begin{align}
    \tilde{\tau}(u) \ket{\Psi_5} &= \Lambda_0(u)^2\ket{\Psi_5} + 2\xi \Lambda_0(u) S^-_L B_R(u) \ket{\Psi_5} \nonumber \\
    &-2\xi \Lambda_0(u)B_L(u) S^-_R \ket{\Psi_5}-4\xi^2B_L(u) S^-_L B_R(u)S^-_R \ket{\Psi_5}.
\end{align}
The term at order $O(\xi^2)$ trivially vanishes 
\begin{align}
     B_R(u)S^-_R \ket{\Psi_5} \propto B_R(u) [S^-_R,S^+_R] \ket{\Omega} \propto B_R(u) \ket{\Omega} = 0. 
\end{align}
Moreover, using the commutation relations~\eqref{eq:commutation-S-B} and~\eqref{eq:commut-A-D-S+}, the contribution at order $O(\xi)$ simplifies to 
\begin{align}
    S^-_L B_R(u) \ket{\Psi_5} &= 2 (J+1) \left((u+1)^J-u^J\right)S^+_L \ket{\Omega}, \\
    B_L(u) S^-_R \ket{\Psi_5} &= 2J\left\{\left((u+1)^J-u^J\right)S^+_L \ket{\Omega} -C_L(u) \ket{\Omega}\right\}.
\end{align}
Expanding $C_L(u)\ket{\Omega}$ in the undeformed one-magnon eigenbasis, we end up with
\begin{align}
     \tilde{\tau}(u) \ket{\Psi_5} = \Lambda_0(u)^2\ket{\Psi_5} - 4\xi \Lambda_0(u)((u+1)^J-u^J)\sum_{k=1}^{J-1} \frac{v_k}{(u-v_k)\left(v_k+1\right)^{J-1}}C_L(v_k) \ket{\Omega}.
\end{align}
Therefore, the state
\begin{align}
    \ket{\tilde{\Psi}_5} = \ket{\Psi_5} + 4\xi \sum_{k=1}^{J-1} \frac{v_k}{\left(v_k+1\right)^{J-1}}C_L(v_k) \ket{\Omega}.
\end{align}
is an eigenstate of $\tilde{\tau}(u)$ with eigenvalue $\Lambda_0(u)^2$.

\vspace{12pt}

In summary, in the subspace of states with a maximum of two excitations of type $L$ and one of type $R$, the twisted transfer matrix~\eqref{eq:deformed-monodromy} at generic values of $u$ is non-diagonalisable, with $J-1$ generalised eigenvectors of rank two, associated in the undeformed limit with the states $\ket{\Psi_3^{(k)}}=C_L(v_k) S^+_L S^+_R \ket{\Omega}$. 

Finally, note that the twisted transfer matrix is invariant under the transformations $L \leftrightarrow R$ and $\xi \to - \xi$. Therefore, the (generalised) eigenvectors in the subspace with at most $n=1$ and $\bar n=2$ excitations are analogous to those in the $n=2$ and $\bar n=1$ case under the formal substitution $L \leftrightarrow R$ and $\xi \to - \xi$.

\section{Spectrum of the twisted transfer matrix in the diagonal form}\label{sec:diagon-Ham}

As discussed in section~\ref{sec:ABA}, in the eigenbasis of the Cartan generators of $\mathfrak{sl}(2)_L \oplus \mathfrak{sl}(2)_R$, the Hilbert space of the model~\eqref{eq:full_Hilbert_space} can be decomposed into subspaces with a fixed maximum number of excitations. Within each subspace, the twisted transfer matrix~\eqref{eq:deformed-monodromy} is represented by an upper triangular, non-diagonalisable matrix that admits a Jordan block decomposition.

However, another strategy is available. Since the model is invariant under the negative-root generators $J^-_N$ (see section~\ref{sec:Moyal-spin-chain}), one can diagonalise the twisted transfer matrix in the common eigenbasis of $J^-_L$ and $J^-_R$. Interestingly, a similar approach has been applied to other models that also arise in deformations of $\mathcal{N}=4$ super Yang--Mills~\cite{Guica:2017mtd,Driezen:2025dww,Driezen:2025izd}.

This strategy is only possible in representations of $\mathfrak{sl}(2)_L \oplus \mathfrak{sl}(2)_R$ with negative spin value. In those cases, the fundamental module is infinite dimensional and one is able to construct explicit eigenstates of the negative-root generators. To be more precise, consider one of the copies of the algebra. In the holomorphic representation~\eqref{eq:hol-rep}, the negative-root generator is the derivative operator in the space of analytic functions,
\begin{align}
    J^-_N = \partial,
\end{align}
where the derivative is taken with respect to the variable $z$ for the $L$ copy, and $\bar z$ for the $R$ copy.
A possible basis of this space is the set of monomials $\{z^n\}$ with $n \in \mathbb{N}$, which is also a basis of the Cartan generator $J^3_N$. In this basis, it is clear that $J^-_N$ is non-diagonalisable, since
\begin{align}
   \left(J^-_N\right)^{n+1} z^n =0, \qquad\qquad J^-_Nz^n=nz^{n-1}. 
\end{align}
In other words, $z^n$ is a generalised eigenvector of rank $n+1$ in the Jordan chain of $z^0=1$ and with eigenvalue $0$. However, since the space is infinite dimensional, one can also consider the basis $\{e^{\lambda z}\}$ with $\lambda$ a real parameter. These states can be understood as infinite power series constructed with eigenstates $z^n$ of the Cartan generator. Then, in this basis
\begin{align}
    J^-_N e^{\lambda z} = \lambda e^{\lambda z},
\end{align}
and the operator is diagonalisable. We have found two bases, one in which the operator is diagonalisable and another in which it is not. These two scenarios would be incompatible if we were working just with matrices, but in this case we are dealing with linear operators on an infinite-dimensional Hilbert space. In fact, the basis $\{z^n,n\in \mathbb N\}$ and the one $\{e^{\lambda z},\lambda\in \mathbb R\}$ are related by the transformation
\begin{align}
    z^n = (-1)^n \int \delta^{(n)}(\lambda) e^{\lambda z} d\lambda, 
\end{align}
where $\delta^{(n)}$ is the $n$-th derivative of the Dirac delta function. The change of basis therefore requires an integration over all the possible eigenstates of the form $e^{\lambda z}$, with coefficients that are distributions.

After discussing the diagonalisability of the negative-root generators in non-compact representations, we aim to diagonalise the twisted transfer  matrix~\eqref{eq:deformed-monodromy} in the common eigenbasis of $J^-_L$ and $J^-_R$. Denote by $\ket{\Psi_{(M_L,M_R)}}$ the simultaneous eigenstates of both global negative-root generators, such that
\begin{align}\label{eq:MLR}
    S^-_N \ket{\Psi_{(M_L,M_R)}} = M_N \ket{\Psi_{(M_L,M_R)}} \quad \text{with} \quad N \in \{L,R\}.
\end{align}
In this basis, the eigenvalue equation for the twisted transfer matrix~\eqref{eq:deformed-monodromy} reads
\begin{align}
    \left(\tau_L(u) -2 \xi M_R B_L(u) \right)\left(\tau_R(u) +2 \xi M_L B_R(u)\right) \ket{\Psi_{(M_L,M_R)}} = \tilde \Lambda(u) \ket{\Psi_{(M_L,M_R)}}.
\end{align}
Notice that in the above equation the operator on the left-hand side factorises into a product of one $L$ and one $R$ operator. In other words, in the basis of eigenstates of the negative-root generators, the twisted transfer matrix decomposes into two deformed $XXX_{-1/2}$ models, one for each copy of the $L$ and $R$ algebra. In fact, defining two effective deformation parameters
\begin{align}
    \xi_L = -\frac{2\xi M_R}{J}, \quad \xi_R=\frac{2 \xi M_L}{J}, \label{eq:effect-param-def}
\end{align}
the eigenvalue equation can be rewritten as
\begin{align}
    \tilde{\tau}_L(u)\tilde{\tau}_R(u) \ket{\Psi_{(M_L,M_R)}} = \tilde\Lambda(u) \ket{\Psi_{(M_L,M_R)}}, \quad \tilde{\tau}_N(u) = \tau_N(u) + \xi_N J B_N(u). \label{eq:dipole-deformed-monodromy}
\end{align}
As a consequence, the eigenstates also factorise into a $L$ and a $R$ component
\begin{align}
    \ket{\Psi_{(M_L,M_R)}} = \ket{\Psi_{(M_L)}} \otimes \ket{\Psi_{(M_R)}}, \label{eq:eigenstate-root}
\end{align}
such that each factor satisfies its own eigenvalue equation
\begin{align}
    \tilde{\tau}_N(u) \ket{\Psi_{(M_N)}} =\tilde\Lambda_N(u) \ket{\Psi_{(M_N)}}.
\end{align}
Interestingly, the twisted transfer matrix $\tilde{\tau}_N(u)$~\eqref{eq:dipole-deformed-monodromy} coincides with the twisted $XXX_{-1/2}$ spin-chain that appears in the dipole-deformed $\mathcal{N}=4$ super Yang--Mills~\cite{Guica:2017mtd}. In particular, it can be obtained via a Drinfel'd twist of the form
\begin{align}
    F_{12} = e^{\frac{i}{2}\xi_N \left(J^-_N \wedge \,\mathbb{I}\right)}.
\end{align}
Therefore, one can compute the spectrum of the Groenewold-Moyal-twisted spin-chain using the methods developed for the dipole deformation. More specifically, the spectrum of the dipole-deformed model was obtained within the Baxter $T-Q$ framework. The starting point is the $T-Q$ relation of the $XXX_{-1/2}$, derived in the context of the separation of variables method~\cite{Derkachov:2002tf}. In our conventions, it reads
\begin{align}
    (u+1)^{J} Q_N(u+1) + u^J Q_N(u-1) = \Lambda_N(u) Q(u)_N, \label{eq:Baxter-eq}
\end{align}
where $\Lambda_N(u)$ is the eigenvalue of the $N$-copy of the undeformed transfer matrix. It was then conjectured in~\cite{Guica:2017mtd} that the dipole deformation holds the same functional equation for the twisted model but with a $Q$-function, $\tilde{Q}_N(u)$, and an eigenvalue $\tilde\Lambda_N(u)$ that depend on the deformation parameter  $\xi_N$. Solving the Baxter equation, one obtains the spectrum through the relation
\begin{align}
    \tilde E = \frac{d}{du} \log\left(\frac{\tilde Q_N(u)}{\tilde Q_N(u-1)}\right)\bigg|_{u=0}. \label{eq:energy-from-Q}
\end{align}
We write down the energy of the vacuum (see eq. (6.28) of~\cite{Guica:2017mtd})
\begin{align}
    \tilde E_{(0)} = \frac{\xi_N^2 M_N^2}{J+1}-\frac{\xi_N^4 M_N^4}{12(J+1)^2}+\frac{(J^2+J+2)\xi_N^6 M_N^6}{360(J+1)^3(J+2)} + O(\xi_N^8 M_N^8), \label{eq:energy-vacuum-dipole}
\end{align}
and the energy of the excited states in a spin-chain of length $J=2$ (see eq. (6.31) of~\cite{Guica:2017mtd})
\begin{align}
     \tilde E_{(j_N)} = 4 h(j_N)- \frac{\xi_N^2 M_N^2}{(2j_N-1)(2j_N+3)} + O(\xi_N^4 M_N^4), \label{eq:energy-excited-dipole}
\end{align}
where $j_N$ is a natural number that labels the irreducible modules of the decomposition $V_F^N \otimes V_F^N$ in~\eqref{eq:VF-VF}. This index coincides with the number of excitations of the lowest-weight state of the irreducible module to which the undeformed limit of the eigenstate belongs.

In our case, for the Groenewold-Moyal-twisted model~\eqref{eq:dipole-deformed-monodromy}, the $Q$-function must factorise as $\tilde Q(u)=\tilde Q_L(u) \tilde Q_R(u)$. Therefore, from~\eqref{eq:energy-from-Q}, it is clear that the energy of the Groenewold-Moyal-deformed $XXX_{-1/2}^{\oplus 2}$  spin-chain, in the eigenbasis of the negative-root generators, coincides with the sum of the energies of two dipole-deformed $XXX_{-1/2}$ spin-chains. In fact, using the identification~\eqref{eq:effect-param-def} and the expression for the energy~\eqref{eq:energy-vacuum-dipole}, one obtains the following energy of the vacuum
\begin{align}\label{eq:en-gr-st}
    \tilde E_{(0)}= \frac{8 \xi^2 M_L^2 M_R^2}{J^2(J+1)}-\frac{8 \xi^4 M_L^4 M_R^4}{3 J^4(J+1)^2}+\frac{16(J^2+J+2)\xi^6 M_L^6 M_R^6}{45 J^6(J+1)^3(J+2)} + O(\xi^8 M_L^8 M_R^8),
\end{align}
while~\eqref{eq:energy-excited-dipole} leads to
\begin{align}
    \tilde E_{(j_L,j_R)} &= 4 \left( h(j_L) + h(j_R)\right) -\left(\frac{1}{(2j_L-1)(2j_L+3)}+\frac{1}{(2j_R-1)(2j_R+3)}\right) \xi^2 M_L^2 M_R^2  \nonumber \\
    &+ O(\xi^4 M_L^4 M_R^4). \label{eq:en-ex-st}
\end{align}
Remarkably, in the eigenbasis of the negative-root generators, the Groenewold-Moyal-deformed Hamiltonian is not only diagonalisable, but also possesses a deformed spectrum. 

Notice that the corrections to the eigenvalues depend on the combination $\xi M_L M_R$.  Therefore, if either $M_L$ or $M_R$ is zero the spectrum is not deformed. Moreover, suppose $M_L=0$ (the same reasoning applies if $M_R=0$). Then, according to~\eqref{eq:dipole-deformed-monodromy}, the $R$-twisted transfer matrix is not deformed and the $R$-eigenvectors coincide with the undeformed ones. However, the $L$-twisted transfer matrix is still deformed and the $L$-eigenvectors receive corrections.  That is to say, in the eigenbasis of the negative root generators $J^-_L$ and $J^-_R$, one can have undeformed eigenvalues associated with non-trivially deformed eigenstates. Finally, if  the eigenstate is a lowest-weight state with respect to both copies of the algebra, i.e. $M_L=M_R=0$, both the spectrum and the eigenfunction remain undeformed.

\subsection{Eigenstates for a spin-chain of length $J=2$}
The exact eigenstates of the twisted model can be obtained by solving the eigenvalue equation~\eqref{eq:dipole-deformed-monodromy}. For simplicity, we will focus on a spin-chain of length $J=2$. In the holomorphic representation~\eqref{eq:hol-rep}, the eigenstates~\eqref{eq:eigenstate-root} of the negative root generators are
\begin{align}
    \ket{\Psi_{(M_L,M_R)}} = e^{\frac{M_L(z_1+z_2)}{2}}e^{\frac{M_R(\bar z_1+\bar z_2)}{2}} h(z) \bar h(\bar z), \quad z=z_2-z_1; \quad \bar z=\bar z_2-\bar z_1
\end{align}
where the variables $z_i$ and $\bar z_i$ are associated with the $L$ and $R$ copies of the algebra, while the index $i \in \{1,2\}$ labels each site of the spin-chain. The functions $h(z)$ and $\bar h(\bar z)$ may be obtained by solving~\eqref{eq:dipole-deformed-monodromy}. In fact, in the holomorphic representation, the twisted transfer matrix is a differential operator on the space of analytic functions, which leads to the following ordinary differential equation for $h(z)$
\begin{align}
    z (z-2 \xi_L) h''(z)+2 (z-\xi_L) h'(z)&+\frac{1}{4} h(z) \left(-M_L^2 z^2+2 \xi_L M_L (M_L z+4 u+2)+ \right. \nonumber \\
   & \left. + 8 u^2+8 u+4\right)  
    =\tilde \Lambda_L(u) h(z) \label{eq:ODE-h}
\end{align}
and an identical one for $\bar h(\bar z)$ under the substitution $z \leftrightarrow \bar z$ and $L \leftrightarrow R$. Notice that~\eqref{eq:ODE-h} must be satisfied for every value of the spectral parameter $u$. This implies that the eigenvalue $\tilde \Lambda_N(u)$ is a polynomial in $u$ of degree two, of the form
\begin{align}
    \tilde \Lambda_N(u) = 2u^2+2\left(1+\xi_L M_L\right)u + \tilde \Lambda_N^{(0)}
\end{align}
In order to obtain the constant $\tilde \Lambda_N^{(0)}$ and the eigenfunctions $h(z)$ and $\bar h (\bar z)$ we need to solve the order $O(u^0)$ of~\eqref{eq:ODE-h}
\begin{align}
    z (z-2 \xi_L) h''(z)+2 (z-\xi_L) h'(z)&+\frac{1}{4} h(z) \left(-M_L^2 z^2+2 \xi_L M_L(M_L z+2)+ 4\right) =\nonumber \\  
    &=\tilde \Lambda_L^{(0)} h(z). \label{eq:ODE0-h}
\end{align}
Doing the change of variables $z=x+\xi_L$ and $\tilde{M}_L=-\frac{i}{2} M_L$,  the differential equation~\eqref{eq:ODE0-h} can be rewritten as
\begin{align}
    (\xi_L^2-x^2)f''(x)-2xf'(x)+\tilde{M}_L^2(\xi_L^2-x^2)f(x) = (t^{(0)}_L+\frac{1}{2})f(x), \label{eq:ODE-dipole}
\end{align}
where we have defined
\begin{align}
    f(x) = h(x+\xi_L), \quad t^{(0)}_L = -\tilde \Lambda^{(0)}_L+2i\xi_L \tilde{M}_L+\frac{1}{2}.
\end{align}
This form of the equation~\eqref{eq:ODE-dipole} coincides with the ODE satisfied by the eigenfunctions of the dipole-twisted $XXX_{-1/2}$ spin-chain (see eq. (4.31) of~\cite{Guica:2017mtd}). The solutions of this equation are the prolate angular spheroidal functions of the first kind. In particular, reversing all changes of variables, we find that in our model the eigenfunctions $h(z)$ and $\bar h(\bar z)$ are
\begin{align}
    h(z) = PS_{j_L,0}\left(\frac{i\xi M_L M_R}{4},-\frac{z}{\xi M_R}\right), \quad \bar h(\bar z)=PS_{j_R,0}\left(-\frac{i\xi M_L M_R}{4},\frac{\bar z}{\xi M_L}\right), \label{eq:deformed-h-w}
\end{align}
while the eigenvalues $\Lambda_N(u)$ of the twisted transfer matrix take the form
\begin{align}
    \tilde \Lambda_L(u) &= 2u^2+2(1-\xi M_L M_R) u +1-\xi M_L M_R+\lambda_{j_L,0}\left(\frac{i\xi M_L M_R}{4}\right), \\
    \tilde \Lambda_R(u) &= 2u^2+2(1+\xi M_L M_R) u +1+\xi M_L M_R+\lambda_{j_R,0}\left(-\frac{i\xi M_L M_R}{4}\right),
\end{align}
where $\lambda_{n,m}(y)$ denotes the spheroidal eigenvalues. 

In the undeformed limit ($\xi \to 0$), the eigenfunctions of the differential equation~\eqref{eq:ODE0-h} reduce to the spherical Bessel functions of the first kind~\cite{Driezen:2025dww,Driezen:2025izd}, 
\begin{align}
    h(z) = BJ\left(j_L,-\frac{i M_L}{2}z\right), \quad  \bar h(\bar z) = BJ\left(j_R,-\frac{i M_R}{2}\bar z\right) \label{eq:undeformed-h-q}
\end{align}
and the eigenvalues take the form
\begin{align}
    \Lambda_N(u) &= 2u^2+2 u +1+j_N(j_N+1), \quad N \in \{L,R\}. \label{eq:undeformed-Lambda}
\end{align} 

Now, suppose $M_L=0$ while $M_R \neq 0$ (the same idea applies in the opposite case). As already explained, in this situation, after the twist, the spectrum~\eqref{eq:undeformed-Lambda} and the eigenfunction $\bar h(\bar z)$  in~\eqref{eq:undeformed-h-q} remain undeformed. However, the $L$-eigenstate is deformed. In fact, in the $M_L=0$ limit the function $h(z)$ in~\eqref{eq:deformed-h-w} reduces to the Legendre polynomials~\cite{Guica:2017mtd}
\begin{align}
    h(z) = P_{j_L}(-\frac{z}{\xi M_R}). \label{eq:Legendre}
\end{align}

These are polynomials of degree $j_L$ in variable $z$, and therefore they are finite linear combinations of eigenstates of $J^3_L$. This should be compared with the situation discussed in section~\ref{sec:ABA}, where we obtained undeformed eigenvalues associated with (generalised) eigenvectors constructed as finite sums of eigenstates of the Cartan generators.

Note that the eigenfunctions~\eqref{eq:Legendre} are singular in the limit $\xi\to 0$ or $M_R \to 0$. However, we are free to choose any normalisation factor. In fact, multiplying~\eqref{eq:Legendre} by 
\begin{equation}
    c_{j_L}=\frac{\left(-2\xi M_R\right)^{j_L}}{\binom{2 j_L}{j_L}}
\end{equation}
gives a function whose leading term reduces to the lowest-weight state with $j_L$ excitations
\begin{align}
     c_n h(z) = z^{j_L}+O( \xi M_R z^{j_L-1}).
\end{align}
In particular, if both $M_L=M_R=0$ the eigenfunctions are lowest-weight states with respect to both copies of the algebra
\begin{align}
    h(z) = z^{j_L}, \quad \bar h(\bar z) = \bar z^{j_R}.
\end{align}
This completes the discussion of the diagonalisation of the twisted transfer matrix~\eqref{eq:deformed-monodromy} in the eigenbasis of the negative root generators.

\section{Match with the string theory side}\label{sec:string}
In this section we want to consider the string-theory side of the AdS/CFT duality, and match a deformed string sigma-model with the spin-chain construction of the previous sections. The deformation of the string sigma-model that we consider falls into the large family of the so-called homogeneous Yang-Baxter deformations~\cite{Klimcik:2002zj,Klimcik:2008eq,Kawaguchi:2014qwa,Matsumoto:2015jja,vanTongeren:2015soa}. Given the Lie algebra of isometries $\mathfrak g=Lie(G)$ of the undeformed model, these deformations are generated by  antisymmetric solutions of the classical Yang-Baxter equation on $\mathfrak{g}$. Knowing that Drinfel'd twists continuously connected to the identity are in one-to-one correspondence with antisymmetric solutions of the classical Yang-Baxter equation~\cite{drinfeld1983constant}, it is natural to expect that the string-theory realisation of the deformation that we consider is via the homogeneous Yang-Baxter construction. We refer to~\cite{Matsumoto:2014gwa,vanTongeren:2015uha} for the papers first exploiting the relation between Drinfel'd twists and solutions of the classical Yang-Baxter equation in the AdS/CFT context. The correspondence between the Drinfel'd twisted spin-chains and the deformed string sigma-models has in fact already been shown to work for other deformations falling within this class, like the $\beta$-deformation~\cite{Frolov:2005iq,Frolov:2005dj}, the dipole deformation of~\cite{Guica:2017mtd}, Jordanian deformations~\cite{Driezen:2025dww,Driezen:2025izd} and the angular dipole deformation of~\cite{Meier:2025tjq}. Here we will confirm this expectation also in our case by working out a non-trivial match.

\subsection{Preliminary setup}
In the previous sections, the spin-chain was twisted with $F_{12}=\exp(\xi J^-_L\wedge J^-_R)$, and expanding the Drinfel'd twist in the deformation parameter $F_{12}=\mathbb I+\xi \ r_{12}+\mathcal O(\xi^2)$ one identifies the $r$-matrix $r_{12}=J^-_L\wedge J^-_R$. At the same time, we can consider a ``dual'' twist obtained by implementing the automorphism of the $\mathfrak{sl}(2)$ algebra $J^+\leftrightarrow J^-, J^3\to -J^3$. In fact, the considerations in section~\ref{sec:diagon-Ham} apply identically also after the automorphism, and the spectrum of the spin-chain is still given by~\eqref{eq:en-gr-st} and~\eqref{eq:en-ex-st}. In the following, we will  prefer to use $r_{12}=J^+_L\wedge J^+_R$  to deform the sigma-model.

For our purposes, it will be enough to consider the bosonic truncation of the string sigma-model, and we will consider $AdS_3\times S^3\times T^4$ as the seed background.\footnote{See later for comments on $AdS_5\times S^5$. The results of this section automatically apply also to the $AdS_5\times S^5$ case, because the deformed sigma model that we will consider can be understood as a consistent truncation of the deformation of $AdS_5\times S^5$.} Given that the $T^4$ factor will only play the role of a spectator, we will really just focus on the $AdS_3\times S^3$ factor. In fact, the deformation will act non-trivially only on the $AdS_3$ factor, so that most of the time we will only talk  about this reduced sigma-model, and we will add the $S^3$ contribution later by hand. The group of isometries of $AdS_3$ is $G=SO(2,2)$, which we will take  to be spanned by the conformal generators: $D$ for scale transformations, $J_{01}$ for the Lorentz boost, $p_\mu$ for translations and $k_\mu$ for special conformal transformations, with $\mu=0,1$. For the conformal algebra we will follow the conventions of~\cite{Borsato:2022ubq}. In particular, we will use the following map to rewrite $\mathfrak g=\mathfrak{so}(2,2)=\mathfrak{sl}(2)_L\oplus \mathfrak{sl}(2)_R$ in terms of the basis used in the spin-chain sections 
\begin{equation}\label{eq:map-sl2-conf}
\begin{aligned}
        & J^+_L=(+p_0+p_1)/\sqrt{2},\qquad && J^-_L=(-k_0+k_1)/(2\sqrt{2}),\qquad &&& J^3_L=(D-J_{01})/2,\\
        & J^+_R=(-p_0+p_1)/\sqrt{2},\qquad && J^-_R=(+k_0+k_1)/(2\sqrt{2}),\qquad &&& J^3_R=(D+J_{01})/2,
\end{aligned}
\end{equation}
so that $r_{12}=J^+_L\wedge J^+_R=p_0\wedge p_1$. A Yang-Baxter deformation with this kind of $r$-matrix gives rise to a Maldacena-Russo-Hashimoto-Itzhaki deformation~\cite{Maldacena:1999mh,Hashimoto:1999ut}, see also~\cite{Alishahiha:1999ci}.\footnote{In this case the deformation involves the time direction. To avoid this while remaining within the class of solutions $r_{12}=p_\mu\wedge p_\nu$, one needs to go to higher dimensional $AdS$.}

The action of the undeformed sigma model can be obtained as a symmetric coset\footnote{The $AdS_3$ sigma-model can be realised also as a Principal Chiral Model on the Lie group $SL(2,\mathbb R)$.}  $G/H$ where $H=SO(1,2)$. One can in fact identify a $\mathbb Z_2$ grading of $\mathfrak{g}$, so that the $\mathfrak{so}(1,2)$ subalgebra spanned by $J_{01}, p_\mu+k_\mu$ has grading 0, while the complement spanned by $D,p_\mu-k_\mu$ has grading 1. We will parameterise the coset element $g\in G/H$  in terms of the coordinates $x^0,x^1,z$ of the Poincar\'e patch, taking $g=\exp(x^\mu p_\mu)\cdot \exp(D\log z )$, because the deformed background looks simple in these coordinates. Notice that the boundary of undeformed $AdS$ is then at $z=0$.
The action $S_a$ of the deformed $AdS_3$ sigma-model can be written as
\begin{equation}
    S_a=-\frac{\sqrt{\lambda}}{4\pi}\int d\tau d\sigma\ {\rm Tr}\left[g^{-1}\partial_+g\ P \mathcal O^{-1}\left(g^{-1}\partial_-g\right)\right],
\end{equation}
where we use light-cone coordinates on the worldsheet $\sigma^\pm=(\tau\pm\sigma)/2$ and $\mathcal O:\mathfrak{g}\to \mathfrak{g}$ is a linear operator on the Lie algebra defined as
\begin{equation}
    \mathcal O=1-\eta {\rm Ad}_g^{-1}\circ R \circ{\rm Ad}_g\circ P.
\end{equation}
When defining $\mathcal O$, we have introduced  a new deformation parameter for the sigma-model that we called $\eta$ and  that later will be related to the deformation parameter $\xi$ of the spin-chain.
Moreover, $P:\mathfrak{g}\to \mathfrak{g}$ is a projector on the subspace of $\mathfrak{g}$ with grading 1, ${\rm Ad}_g$ denotes the adjoint action so that for $x\in \mathfrak{g}$  we have ${\rm Ad}_gx=gxg^{-1}$, and finally $R:\mathfrak{g}\to \mathfrak{g}$ is a solution of the classical Yang-Baxter equation written as
\begin{equation}
    [R(x),R(y)]-R([R(x),y]+[x,R(y)])=0,\qquad x,y\in \mathfrak{g},
\end{equation}
which is antisymmetric with respect to the trace ${\rm Tr}[xR(y)]=-{\rm Tr}[R(x)y]$. This $R$-matrix is related to the previous $r$-matrix as  $R(x)={\rm Tr}_2(rx)$, where ${\rm Tr}_2$ denotes the trace in the second factor of $r_{12}$.
These ingredients are enough to conclude that the deformed model is still integrable even in the presence of the deformation, where integrability is ensured by a Lax connection satisfying $\partial_+\mathcal L_--\partial_-\mathcal L_++[\mathcal L_+,\mathcal L_-]=0$. Explicitly, the Lax is given by
\begin{equation}\label{eq:Lax}
    \mathcal L_\pm=A^{(0)}_\pm+\zeta^{\mp 1}A^{(1)}_\pm,\qquad
    A_\pm=\frac{1}{1\pm \eta {\rm Ad}_g^{-1}\circ R \circ{\rm Ad}_g\circ P}(g^{-1}\partial_\pm g),
\end{equation}
where $\zeta$ is the spectral parameter, and the superscripts indicate the projections on the subspaces of grading 0 or 1.

The sigma-model action $S_s$ on $S^3$ can be constructed similarly, and in our case it will remain undeformed. The full action $S=S_a+S_s$ of  the deformed sigma-model can also be rewritten in terms of a background metric and $B$-field as
\begin{equation}
    S=-\frac{\sqrt{\lambda}}{4\pi}\int d\tau d\sigma\ (\tilde G_{MN}+\tilde B_{MN})\partial_+X^M\partial_-X^N,
\end{equation}
where $X^M$ are the coordinates parameterising the full 6-dimensional manifold that is a deformation of $AdS_3\times S^3$. Moreover, $\sqrt{\lambda}$ is proportional to the string tension, and in fact $\lambda$ is identified with the 't Hooft coupling in AdS/CFT. Denoting by $X^m=\{x^0,x^1,z\}$ with $m=0,1,2$ the coordinates for the (deformed) AdS, in our case a quick alternative way to obtain the deformed background $(\tilde G_{mn},\tilde B_{mn})$ in terms of the undeformed one  $(G_{mn},B_{mn})$ is given by the formula~\cite{Araujo:2017jkb,Araujo:2017jap,Bakhmatov:2017joy,Borsato:2018idb}
\begin{equation}
    \tilde G+\tilde B = (G+B)\left(1-\eta \Theta(G+B)\right)^{-1},
\end{equation}
where
\begin{equation}
    \Theta= \begin{pmatrix}
        0&1&0\\
        -1&0&0\\
        0&0&0
    \end{pmatrix}.
\end{equation}
Given that the undeformed $AdS_3$ metric is just $G=\text{diag}(-z^{-2},z^{-2},z^{-2})$ and that the undeformed $B$-field vanishes $(B=0)$, using the above formula we find the deformed metric and $B$-field\footnote{When deforming $AdS_5$, the deformed metric has the additional piece $((dx^2)^2+(dx^3)^2)/z^2$ that remains undeformed, and the $B$-field stays as in the $AdS_3$ case. In the case of $AdS_5$ one has more options to implement deformations of this kind, and in particular one can also deform via an $r$-matrix with only spacelike momenta $r_{12}=p_1\wedge p_2$. The previous construction can be easily generalised by taking $\Theta$ to be non-trivial along those directions.}
\begin{equation}
\begin{aligned}
    &d\tilde s^2_a=\tilde G_{mn}dX^mdX^n = \frac{z^2\, dx^\mu dx_\mu}{z^4-\eta^2}+\frac{dz^2}{z^2},\\
    &\tilde B=\frac{1}{2}\tilde B_{mn}dX^m\wedge dX^n = -\frac{\eta}{z^4-\eta^2}dx^0\wedge dx^1.
\end{aligned}
\end{equation}
Obviously, the above metric for the deformed AdS should be accompanied by the one for the undeformed sphere.
The $r$-matrix that we are considering is abelian, and the deformed model can be equivalently obtained by implementing a TsT deformation along $x^0,x^1$~\cite{Osten:2016dvf}. When embedding this classical bosonic sigma-model into string theory, there is also a dilaton arising from the deformation and a transformation of the RR fluxes, but we will not need this information for our purposes.

\subsection{Comments on the spectral problem}
The above deformation breaks most of the original $SO(2,2)$ isometries, and the only generators of this group that survive are $p_0,p_1,J_{01}$.\footnote{This can be easily obtained for example by noticing that the surviving isometries $t\in \mathfrak{g}$ are those whose adjoint action commutes with the $R$-operator, so that ${\rm ad}_tR=R\, {\rm ad}_t$. In the $AdS_5$ case, also the generators $p_2,p_3,J_{23}$ survive in the deformation.} The situation is therefore quite different compared with the undeformed case, where the spectral problem is defined by picking two Cartans in $SO(2,2)$. For example, with no deformation one can choose $p_0-k_0$ and $p_1+k_1$, the former being the generator of global time translation in AdS. The advantage of doing this is that, in particular, one has a timelike Killing vector defined everywhere---something which is easier to see in global coordinates. The conserved charge  corresponding to the symmetry for the global time translations is the energy of the string configurations, and in the undeformed $AdS_5/CFT_4$ setup, for example,  where we have a complete understanding of the AdS/CFT dictionary, this energy is dual to scaling dimensions of the CFT.

In our case, if we wanted to define the spectral problem using only the residual isometries, we would be forced to pick $p_0,p_1$ as possible commuting charges. Notice that these are not Cartan generators because their adjoint action is not diagonalisable. Nevertheless, they may give rise to a meaningful spectral problem if one can organise the string sigma-model solutions in terms of eigenstates of these generators. To make a comparison, in the case of the dipole~\cite{Guica:2017mtd} and Jordanian~\cite{Borsato:2022drc,Driezen:2025dww,Driezen:2025izd} deformations $p_0-k_0$ is also broken, but there is nevertheless a residual isometry corresponding to a new generator of global time translations; this  is now a linear combination of $p_0-k_0$ and $p_1+k_1$, and therefore is still Cartan. At the same time, both in the dipole and Jordanian  cases, a lightlike momentum (e.g.~$p_0+p_1$) is used to label states in the spectral problem, and that generator is certainly not Cartan (its adjoint action is not diagonalisable). It seems, therefore, that one may relax the requirement of having Cartan generators to identify the spectral  problem.\footnote{The need to go beyond the restriction to Cartan generators when considering the classical spectral curve appears also in the context of non-relativistic strings~\cite{Fontanella:2022wfj,Fontanella:2026gaq}.} This is reminiscent of the fact that in the undeformed case one can study AdS spinning strings~\cite{Giombi:2009gd}. However, these massless AdS solutions do not fix a relation between  AdS charges and the angular momentum in the sphere; for this reason, we do not expect that they are related to the spin-chain construction that we have in the previous sections.

To get some clues about the correct identification of the spectral problem in the presence of this deformation, we will explicitly construct a pointlike string solution that is a generalisation of the one of BMN~\cite{Berenstein:2002jq}, which may be understood as the vacuum of the spectral problem of undeformed AdS/CFT integrability. We will then match a conserved charge for this string solution with the ground state energy of the spin-chain.

\subsection{A classical BMN-like solution}
In the undeformed setting, the BMN solution is a pointlike string, so that the embedding coordinates $X^M$ only depend on worldsheet time $\tau$ and do not depend on worldsheet space $\sigma$. If $t$ is the coordinate for global time in AdS and $\phi$ is the angle parameterising a big circle in the sphere, then the solution is obtained by taking $t=\kappa \tau, \phi=\omega\tau$. The parameter $\kappa$ therefore relates target-space and worldsheet times, and $\omega$ is then the angular velocity on the sphere. The two parameters are related to target-space charges as $E=\sqrt{\lambda}\kappa$ and $J=\sqrt{\lambda}\omega$, where $E$ is the energy and $J$ is the angular momentum. The Virasoro constraints on the solution read $-\dot t^2+\dot \phi^2=-\kappa^2+\omega^2=0$, and choosing for example the solution $\kappa=\omega$ fixes a simple relation between the energy and the angular momentum, $E=J$. Our goal now is to generalise this story in the presence of the deformation. 

When taking the undeformed limit of our construction, we will actually consider a slight generalisation of the BMN solution. In Poincar\'e coordinates we have\footnote{In~\cite{Dobashi:2002ar,Tsuji:2006zn} a different type of ``tunnelling solution'' was argued to be relevant for the holographic description. This tunnelling solution may be understood as the analytic continuation of worldsheet time $\tau\to -i\tau$, so that $z\propto 1/\cosh(\kappa\tau)$ and the solution starts and ends at the boundary $z=0$ respectively when $\tau=-\infty$ or $\tau=+\infty$. Here we are simply taking the BMN solution written in global coordinates and rewriting it in Poincar\'e coordinates, inverting the relations $x^0/z=\cosh \rho \sin t, x^1/z=\cos\zeta \sinh\rho, (1+z^2-(x^0)^2+(x^1)^2)/(2z)=\cosh\rho\cos t$, see e.g.~\cite{Tseytlin:2010jv}.}
\begin{equation}
\begin{aligned}
  &      x^0(\tau)=\frac{\alpha^++\alpha^-}{2}\tan(\kappa\tau),\qquad
   &&     z(\tau)=\frac{\sqrt{\alpha^+\alpha^-}}{\cos(\kappa\tau)},\\
&    x^1(\tau)=\frac{\alpha^+-\alpha^-}{2}\tan(\kappa\tau),\qquad
&&    \phi(\tau)=\omega\tau,
\end{aligned}
\end{equation}
so that the standard BMN may be recovered by taking $\alpha^+=\alpha^-=1$. Here we need to generalise it because we want to have the most general charges associated with translations of $x^\mu$. These may be calculated by $q_\mu=K_\mu^mG_{mn}\dot X^n$  where we use the Killing vectors $K_\mu=\partial_\mu$. Importantly, they allow us to have general charges for $L$ and $R$ translations associated with the generators $J^+_L,J^+_R$
\begin{equation}
    q_L=\frac{q_0+q_1}{\sqrt{2}}=-\frac{\kappa}{\sqrt{2}\alpha^+},\qquad\qquad
    q_R=\frac{-q_0+q_1}{\sqrt{2}}=\frac{\kappa}{\sqrt{2}\alpha^-}.
\end{equation}
We want $q_L,q_R$ to be general because they play precisely the same role as the $M_L,M_R$  eigenvalues in the spin-chain formulation, see~\eqref{eq:MLR}, and later they will be identified.

To construct a generalisation of the above solution that is valid in the deformed background, we will solve the equations of motion and the Virasoro constraints when $\eta\neq 0$, and when assuming that $x^0(\tau),x^1(\tau),z(\tau)$ are time-dependent but still do not depend on the spatial coordinate  $\sigma$. We will of course still take $\phi(\tau)=\omega\tau$. The equations of motion can be derived directly from the action and are equivalent to the geodesic equation $\ddot{X}^m+\tilde \Gamma^m_{np}\dot{X}^n\dot{X}^p=0$, with $\tilde \Gamma^m_{np}$ the Christoffel symbols of the deformed metric. The geodesic equation yields a system of three coupled differential equations in the three different functions $x^0(\tau),x^1(\tau),z(\tau)$ that may seem difficult to solve, but we can integrate two of these equations by noticing that also in the presence of the deformation we still have shift isometries with Killing vectors $K_\mu=\partial_\mu$. Also when $\eta\neq 0$, we continue to call 
\begin{equation}
    q_\mu=K_\mu^m\tilde G_{mn}\dot X^n=\frac{z^2}{z^4-\eta^2}\dot{x}_\mu,\qquad \mu=0,1,
\end{equation}
the corresponding conserved charges, and inverting these two relations we can substitute $\dot x^\mu$ into the equations of motion. The only equation left is then
\begin{equation}\label{eq:eomsm}
    z\ddot z-\dot z^2+2q_{L}q_R(z^4+\eta^2)=0.
\end{equation}
At the same time, we must solve the Virasoro constraints, which are given by $\tilde G_{MN}\dot{X}^M\dot{X}^N=\tilde G_{mn}\dot{X}^m\dot{X}^n+\omega^2=0$, and explicitly read
\begin{equation}\label{eq:vira}
    \dot z^2+2q_{L}q_R(z^4-\eta^2)+\omega^2z^2=0.
\end{equation}
It turns out that ~\eqref{eq:eomsm} and~\eqref{eq:vira}  are solved by a Jacobi elliptic function
\begin{equation}\label{eq:sol-z}
    z(\tau)=\frac{z_0}{{\rm cn}(\kappa \tau|m)},
\end{equation}
where\footnote{To check that this solves the equations, one needs  the identities ${\rm dn}^2=1-m \, {\rm sn}^2, {\rm sn}^2+{\rm cn}^2=1$.}
\begin{equation}\label{eq:defBMN}
\begin{aligned}
    &\kappa=\omega\left(1+16\eta^2\omega^{-4}q_L^2q_R^2\right)^{1/4},\qquad
    m=\frac12\left(1-\frac{\omega^2}{\kappa^2}\right),\\
    &z_0=\frac12\sqrt{-\frac{1}{q_Lq_R}\left(\omega^2+\kappa^2\right)}.
    \end{aligned}
\end{equation}
Using this result, one can also integrate the remaining two equations and obtain
\begin{equation}\label{eq:sol-x01}
    x_\mu(\tau)=\frac{q_\mu \left(\text{cn}(\kappa  \tau |m) \left(\tau  \left(\kappa ^2+\omega ^2\right)-2 \kappa  \mathcal{E}(\kappa  \tau |m)\right)+\kappa  \text{dn}(\kappa
    \tau |m) \text{sn}(\kappa  \tau |m)\right)}{4 q_L q_R \text{cn}(\kappa  \tau |m)},
\end{equation}
where $\mathcal{E}(\kappa  \tau |m)$ is the Jacobi epsilon function.

\subsection{Match of the conserved charges via integrability}
At this point we want to use the above classical string solution to identify a conserved charge that is dual to the spin-chain Hamiltonian, i.e.~playing the role that  the energy plays in the undeformed limit.
We will start from the observation that, already in the undeformed setup,  it is possible to match spin-chain and sigma-model calculations in the large-$J$ limit, where in the spin-chain $J$ has the interpretation of length of the chain, while in the string sigma-model $J$ is an angular momentum in the sphere~\cite{Beisert:2003ea,Kruczenski:2003gt,Stefanski:2004cw}. Our aim is therefore to compute a conserved charge on the previous classical string solution and match its large-$J$ expansion with that of the ground state energy of the spin-chain.

In fact, the same strategy was used also for the dipole~\cite{Guica:2017mtd} and Jordanian~\cite{Driezen:2025dww,Driezen:2025izd} deformations, although there the situation is simpler compared to our setup. Crucially, in those cases  it was possible to identify global coordinates for the deformed background ensuring that shifts of a global time $T$ are still isometries. In particular, in those cases on the classical solution one still has $T=\kappa\tau$,  the energy is still given by $E=\sqrt{\lambda}\kappa$, and the angular momentum in the sphere is still $J=\sqrt{\lambda}\omega$. For both the dipole and Jordanian deformations, there is actually another charge, that we may call $M$, that is the eigenvalue of a non-Cartan charge and that plays an important role in the spectral problem, because the Virasoro constraint fixes $E=\sqrt{J^2+\eta^2 M^2}$. It was then  shown that by taking the large-$J$ limit it was possible to match the string-theory results with the ones coming from the spin-chains. In fact, taking into account that $E=J+\frac{\eta^2}{2} M^2\ J^{-1}+\mathcal O(J^{-3})$, they were able to reproduce the $\mathcal O(J^{-1})$ contribution in the large-$J$ expansion of the energy of the spin-chain ground state. Remarkably, the authors of~\cite{Driezen:2025izd} were able to match even the next-to-leading term by considering semiclassical fluctuations of the integrability spectral curve.

In our case we can still introduce charges
\begin{equation}
    J=\sqrt{\lambda}\omega,\qquad Q_L=\sqrt{\lambda}q_L,\qquad Q_R=\sqrt{\lambda}q_R,
\end{equation}
related to the isometries surviving the deformation. We will identify the angular momentum $J$ with the spin-chain length as in the undeformed case. We will also identify the left/right charges of the string sigma model and of the spin-chain as
\begin{equation}
    Q_L=M_L,\qquad Q_R=M_R,
\end{equation}
because they are charges for the same generators $J^+_L,J^+_R$ on the two sides of the AdS/CFT duality.

The challenge now is to identify a conserved charge whose large-$J$ expansion matches  that of the energy of the ground state of the spin-chain calculated in~\eqref{eq:en-gr-st}
\begin{equation}\label{eq:expansion-E0}
    \frac{\lambda}{8\pi^2}\tilde E_{(0)}=\frac{\lambda\xi^2}{\pi^2}\frac{M_L^2M_R^2}{J^3}+\mathcal O(J^{-4}),
\end{equation}
where we rescaled the spin-chain Hamiltonian by $\frac{\lambda}{8\pi^2}$ as one does in the undeformed case~\cite{Beisert:2003jj}, where $\lambda$ is the 't Hooft coupling. The first interesting observation is that both the spin-chain result in~\eqref{eq:en-gr-st},~\eqref{eq:expansion-E0} and the string-theory result in~\eqref{eq:sol-z},~\eqref{eq:defBMN} and~\eqref{eq:sol-x01} only depend on the product $M_LM_R$, and not on left and right charges separately. This is  a non-trivial compatibility requirement that is already satisfied! 

While for dipole and Jordanian deformations the relevant term in the large-$J$ expansion was $J^{-1}$, from~\eqref{eq:expansion-E0} we see that in our case we should expect $J^{-3}$. To reproduce the $J^{-3}$ terms of the spin-chain, it would be tempting to employ a naive generalisation of  the formulas of the undeformed, dipole and Jordanian cases because
\begin{equation}
    \sqrt{\lambda}\kappa = J\left(1+16\eta^2J^{-4}M_L^2M_R^2\right)^{1/4}= J+4\eta^2\frac{M_L^2M_R^2}{J^3}+\mathcal O(J^{-4}),
\end{equation}
which would lead to the identification of the two deformation parameters of the sigma-model and the spin-chain as
\begin{equation}\label{eq:rel-eta-xi}
    \eta=\frac{\sqrt{\lambda}}{2\pi}\xi.
\end{equation}
Although this match is highly non-trivial and encouraging, it feels unsatisfactory and it calls for a stronger justification of the identification of the charge. In fact, while in the previous cases $E=\sqrt{\lambda}\kappa$, in our case we do not have an interpretation for $\sqrt{\lambda}\kappa$ as a conserved charge. 

To have a more complete picture, one may  try finding adapted coordinates so that the above pointlike solution has target-space fields that are linear in $\tau$ in the new coordinate system, like in the dipole and Jordanian deformations, in the hope to find $T=\kappa\tau$. Looking for this new coordinate system, though, would make sense only if the map between the Poincar\'e and the new coordinates is independent of the parameters $\omega,q_Lq_R$ of the solution. Given a massive geodesic in the deformed AdS (which is the case we are considering), the construction of Fermi normal coordinates~\cite{Manasse:1963zz} indeed allows one to find a new coordinate system so that on the geodesic the target-space time is  $T\propto\tau$, and the transverse coordinates are just 0. This logic, however, has two main issues. First, Fermi normal coordinates do not fix the proportionality coefficient between $T$ and $\tau$, simply because one can always perform a coordinate redefinition to reabsorb it; having this extra freedom is a downside  for our purposes. Second, after identifying the new time coordinate $T$ with the procedure of Fermi normal coordinates, the metric is not expected to be invariant under shifts of $T$, because we already know that the only isometries left are $p_0,p_1,J_{01}$. This strategy of Fermi normal coordinates, then, does not seem useful to identify a  conserved charge.

We will instead exploit the classical integrability of the sigma-model under study and compute the monodromy matrix, which we will then use to identify a suitable conserved charge proposed to be dual to the spin-chain Hamiltonian. Given the Lax in~\eqref{eq:Lax}, the monodromy matrix is given by
\begin{equation}
    \Omega=\mathcal P\exp\left(\int_0^{2\pi} d\sigma' \ \mathcal L_\sigma(\tau,\sigma',\zeta)\right),
\end{equation}
where $\mathcal P\exp$ is the path-ordered exponential. Since we will evaluate $\Omega$ on the previous classical solution, the Lax will actually be independent of $\sigma$ and  we can just write
\begin{equation}
    \Omega=\exp\left(2\pi \mathcal L_\sigma(\tau,\zeta)\right).
\end{equation}
Explicitly, on the classical solution we find
\begin{equation}
          \mathcal L_\sigma(\tau,\zeta)=\alpha^\mu p_\mu+\beta^\mu k_\mu+\gamma D,
\end{equation}
with
\begin{equation}
\begin{aligned}
   \alpha^0=&-\frac{(\zeta +1)  \left((\zeta +1) \eta  q_1+(1-\zeta )q_0 z^2\right)}{4 \zeta  z},\quad
   \beta^0=\frac{(\zeta -1)  \left((\zeta -1) \eta  q_1-(\zeta +1) q_0 z^2\right)}{4 \zeta  z}\\
   \alpha^1=&+\frac{(\zeta +1)
    \left((\zeta +1) \eta  q_0+(1-\zeta  )q_1 z^2\right)}{4 \zeta  z},\quad
     \beta^1=\frac{(\zeta -1) \left((1-\zeta ) \eta  q_0+(\zeta +1) q_1 z^2\right)}{4 \zeta  z}\\
\gamma=&    -\frac{\left(\zeta ^2-1\right) \dot z}{2 \zeta  z} .
\end{aligned}
\end{equation}
Notice that the Lax and the monodromy matrix are time-dependent because they depend on $z(\tau)$. However, their eigenvalues are constant in time, because they encode the tower of integrability charges evaluated on the classical solution. Indeed, one may check that we can diagonalise the Lax if we take
\begin{equation}
    h\ \mathcal L_\sigma(\tau,\zeta)\ h^{-1} = \lambda_0(p_0-k_0)+\lambda_1(p_1+k_1),
\end{equation}
with
\begin{equation}
    h=\exp\left((b_++b_-)k_0+(b_+-b_-)k_1\right)\cdot \exp\left((a_++a_-)p_0+(a_+-a_-)p_1\right),
\end{equation}
and\footnote{Here we made some choices regarding signs in front of the square roots. Some of these signs are inconsequential, while others were needed to match with the conventions in the undeformed limit.}
\begin{equation}
    \begin{aligned}
        a_\pm&= \frac{- \gamma+\sqrt{ \gamma ^2-2 (\beta^0\mp\beta^1) \left(\sqrt{-4 (\alpha^0\pm\alpha^1) (\beta^0\mp\beta^1)- \gamma ^2}-2 \alpha^0\mp2 \alpha^1\right)} }{4 (\beta^0\mp\beta^1)},\\
   b_\pm&=\frac{\sqrt{ \gamma ^2-2 (\beta^0\pm\beta^1) \left(\sqrt{-4 (\alpha^0\mp\alpha^1) (\beta^0\pm\beta^1)-\gamma  ^2}-2 \alpha^0\pm2 \alpha^1\right)}}{2 \sqrt{-4 (\alpha^0\mp\alpha^1) (\beta^0\pm\beta^1)- \gamma ^2}}.
    \end{aligned}
\end{equation}
Then the eigenvalues are
\begin{equation}
    \lambda_0=\frac{1}{4}(\lambda_++\lambda_-),\qquad
    \lambda_1=\frac{1}{4}(\lambda_+-\lambda_-),
\end{equation}
with
\begin{equation}
    \begin{aligned}
       \lambda_\pm&=\sqrt{-4 (\alpha^0\pm\alpha^1) (\beta^0\mp\beta^1)- \gamma ^2}
=\frac{1}{2}\sqrt{\frac{\left(\zeta ^2-1\right) \left(\left(\zeta ^2-1\right) \omega ^2\mp8 \zeta  \eta  q_Lq_R\right)}{\zeta ^2}},
    \end{aligned}
\end{equation}
which are indeed constant in time.
First of all, we recall  that in the undeformed limit  global charges are obtained when expanding around $\zeta=1$. Also here, after taking $\eta\to 0$, one obtains
\begin{equation}
    \sqrt{\lambda}\lambda_0=J\frac{\epsilon}{2}+\mathcal{O}(\epsilon^2),\qquad\text{when }\zeta=1+\epsilon.
\end{equation}
The point $\zeta=1$ is in fact important for the definition of the undeformed spectral problem, because  (the Cartan subalgebra of) the global charges identified at $\zeta=1$ serve for the labelling of the states and for the identification of the energy.
It turns out that it does not make sense to continue to do this in the presence of the deformation, because if we still try to expand around $\zeta=1$ when $\eta\neq 0$, we find
\begin{equation}
    \sqrt{\lambda}\lambda_0=\sqrt{\frac{i}{2}\eta q_Lq_R}\ \sqrt{\epsilon}+\mathcal O(\epsilon^{3/2}).
\end{equation}
The first observation is that there is a problem of order of limits, because the $\eta\to 0$ limit of the above relation does not reproduce the previous one. Moreover, the square-root of the (shifted) spectral parameter appears, which is related to the non-diagonalisability of the twist under study, see~\cite{Borsato:2021fuy} for comments.

We therefore learn that  in order to find a charge dual to the Hamiltonian of the spin-chain we should expand around a different value of $\zeta$. The naive guess is that this special point should be given by a function of the deformation parameter $\eta$ like $\zeta=1+a_1\eta+a_2\eta^2+\ldots$, so that when sending $\eta\to 0$ we go back to the spectral problem definition of the undeformed case. But we will soon see that in fact it is not the case.
Let us call $\Lambda$ the charge that should be dual to the spin-chain Hamiltonian, and for which we require that
\begin{equation}
    \Lambda = J+\frac{\lambda\xi^2}{\pi^2}\frac{M_L^2M_R^2}{J^3}+\mathcal O(J^{-4}).
\end{equation}
For the match to work, we will also assume that the two deformation parameters are related by direct proportionality $\eta\propto \xi$.
If we define $\Lambda$ in terms of the eigenvalue $\lambda_0$ as
\begin{equation}
    \Lambda=\sqrt{\lambda} \frac{4 \zeta}{ \zeta^2-1}\lambda_0,
\end{equation}
with a proper overall factor to correctly normalise the leading coefficient, we find\footnote{Here we are assuming that we can take $\frac{\zeta}{\zeta^2-1}=\sqrt{\frac{\zeta^2}{(\zeta^2-1)^2}}$, which is true when $\text{Im}(\zeta)\geq0$ and $|\zeta|\geq1$, or when $\text{Im}(\zeta)\leq0$ and $|\zeta|\leq1$.}
\begin{equation}
    \Lambda = J-8\eta^2\frac{\zeta^2}{(\zeta^2-1)^2}\frac{M_L^2M_R^2}{J^3}+\mathcal O(J^{-4}).
\end{equation}
It is clear that it is possible to match the $\mathcal{O}(J^{-3})$ coefficient, but  expanding around $\zeta=1+a_1\eta+a_2\eta^2+\ldots$ is not useful because it would mess up the $\eta$-dependence. 
Instead, assuming that $\zeta$ is fixed, to match the $\mathcal{O}(J^{-3})$ coefficient it is enough to identify the deformation parameters as
\begin{equation}\label{eq:gen-rel-eta-xi}
    \eta=i\frac{1- \zeta^2}{ \zeta\sqrt{2}}\frac{\sqrt{\lambda}}{2\pi}\xi.
\end{equation}
Notice that the proportionality factor $i\frac{1- \zeta^2}{ \zeta\sqrt{2}}$ must still be real, which means that either $\zeta$ is purely imaginary, or that it lies on the unit circle. Interestingly, when using~\eqref{eq:gen-rel-eta-xi} the $ \zeta$-dependence in $\Lambda$ drops and we have\footnote{As a side remark, let us mention that the other eigenvalue of the Lax reads as
\begin{equation}
    \sqrt{\lambda}\lambda_1= \frac{\zeta^2-1}{8\zeta} \left(\sqrt{J^2+i \xi \frac{2  \sqrt{2} \sqrt{\lambda } M_LM_R  }{\pi }}-\sqrt{J^2-i \xi \frac{2  \sqrt{2} \sqrt{\lambda } M_LM_R}{\pi }}\right)=\frac{i \left(\zeta ^2-1\right) \sqrt{\lambda}  M_LM_R \xi }{2 \sqrt{2} \pi  \zeta  J}+\mathcal O(J^{-5}).
\end{equation}
}
\begin{equation}
\begin{aligned}
    \Lambda&=\frac{1}{2} \left(\sqrt{J^2+i \xi \frac{2  \sqrt{2} \sqrt{\lambda } M_LM_R  }{\pi }}+\sqrt{J^2-i \xi \frac{2  \sqrt{2} \sqrt{\lambda } M_LM_R}{\pi }}\right)\\
    &=\left(J^4+\frac{8 \lambda \xi ^2 M_L^2M_R^2 }{\pi ^2}\right)^{1/4} \cos \left(\frac{1}{2} \arctan\left(\frac{2 \sqrt{2} \sqrt{\lambda }\xi M_LM_R  }{\pi 
   J^2}\right)\right).
\end{aligned}
    \end{equation}
We see that this $\Lambda$, although quite similar, is not the previous guess of $\sqrt{\lambda}\kappa$, and in fact the extra factor with the cosine of the arctangent starts contributing at $\mathcal O(J^{-4})$. 

Let us now comment on the $\zeta$-dependence in the identification of the deformation parameters. The interpretation is that in general we have some freedom for the proportionality factor between $\eta$ and $\xi$, and different choices of the proportionality factor correspond to  different choices for the point  in the $\zeta$-plane  where the eigenvalue of the Lax is evaluated. Notice that when $\zeta=i^{1/2}$ then $\eta=\frac{\sqrt{\lambda}}{2\pi}\xi$ as in~\eqref{eq:rel-eta-xi}.
It would be interesting to understand if this freedom in the value of $\zeta$ is only an artefact of the calculation we are doing, and if the value of $\zeta$ can be fixed by considering higher orders in the large-$J$ expansion or by matching excited states.

Let us stress that the conserved charge identified by evaluating the (eigenvalues of the) monodromy matrix at $\zeta\neq 1$ will be non-local; in fact, in the undeformed setup the locality of the corresponding charges is a direct consequence of expanding around a zero of the (gauge transformed) Lax. Moreover, local charges are those corresponding to the residual isometries, and $\Lambda$ is certainly not a charge for them.
We also add that the fact that the value of $\zeta$ at which we expand the monodromy matrix is $\eta$-independent is not so surprising. In fact, already when $\eta\to 0$ we should not expect to recover the usual description of the spectral problem because in this case, instead of labelling states only with Cartan generators, we allow ourselves to use also $J^+_L\propto p_0+p_1,J^+_R\propto p_0-p_1$. Given that these generators do not commute with the usual BMN time generator $p_0-k_0$, already in the undeformed limit we must look for an alternative conserved charge to define the spectral problem, and this must necessarily be non-local because all local options have already been exhausted.

\section{Conclusions}\label{sec:conclusions}
In this paper we took the first steps to apply the methods of integrability to the spectral problem of AdS/CFT dual pairs deformed by Groenewold-Moyal twists. The twisted spin-chain that we constructed here can be understood as a deformation of a subsector of the $AdS_3/CFT_2$ spin-chain for strings on $AdS_3\times S^3\times T^4$. In the $AdS_5/CFT_4$ spin-chain there is no closed $\mathfrak{sl}(2)^2$ sector, not even at one loop, which means that the construction of our paper must be modified to make it relevant for that case. In particular, looking for two commuting generators both corresponding to positive (or to negative) roots, a minimal choice to embed the Groenewold-Moyal twist is to consider an $\mathfrak{sl}(3)$ subsector.
At the same time, the construction and the considerations in section~\ref{sec:string} are automatically valid also in the case of $AdS_5\times S^5$, because in that section it is enough to consider a reduced sigma-model that is a consistent truncation of $AdS_n\times \mathcal M$ spaces, with $n\geq 3$ and $\mathcal{M}$ a compact manifold. We therefore expect some of our results to be valid also for the Groenewold-Moyal deformation of the $\mathcal N=4$ super Yang-Mills spin-chain; for example, the fact that at large $J$ the ground state energy will go like $\xi M_LM_R J^{-3}$ should be true also in that case.

%Something interesting about our results for the twisted spin-chain is that they may be obtained by putting together building blocks of previous calculations for the dipole deformation~\cite{Guica:2017mtd}, at least to one loop, i.e.~nearest-neighbour approximation for the spin-chain Hamiltonian.

Something that the Groenewold-Moyal, the dipole and the Jordanian twists have in common is that they all break at least some of the original Cartan generators that in the undeformed limit are used to label the states in the description of the spectral problem. This forces us to rethink the whole setup and identify different labels to diagonalise the Hamiltonian. In all these cases, some non-diagonal generators entering the definition of the twist ($J^+_L,J^+_R$ in the case of the present paper) appear to be the necessary ones to provide new labels for the states of the spectral problem. In other words, when attempting to diagonalise the Hamiltonian, we should look for mutual eigenstates of the Hamiltonian and of these non-diagonal generators. This point of view is quite drastic, in the sense that even when taking the undeformed limit one does not recover the usual spectral problem description, because the basis of the Hilbert space has been reorganised in terms of eigenstates of different non-diagonal charges, rather than of the usual Cartans.  In this respect, it seems natural to try to understand the spectral problem already of the undeformed integrable model in the new basis, as it will be the starting point to understand the one for the twisted models.

In fact, in this paper we showed that also for the Groenewold-Moyal twist deformation, as for the other non-diagonal ones, if one insists on working with a basis formed by \emph{finite} linear combinations of eigenstates of the Cartans, then the twisted Hamiltonian appears to be of Jordan block form, in general. In fact, in section~\ref{sec:ABA} we demonstrated effective methods to explicitly calculate the eigenstates of the Hamiltonian, when they exist, and generalised eigenstates more generally. Despite the non-diagonalisability in this picture, it may still be useful to work in this basis because  the (generalised) eigenvalues remain undeformed and can therefore be calculated with the usual Bethe equations; moreover, the Hamiltonian does take a simple form, albeit not diagonal.  Given that in the undeformed case the spin-chain eigenstates correspond to those operators diagonalising 2-point functions of the holographic CFT, it is worth exploring if the basis in which the Hamiltonian is of Jordan-block form is also useful to have a simple understanding of correlation functions of the twist-deformed gauge theory.

The results of this paper may actually help clarify an additional issue present in the case of the Groenewold-Moyal twist, and that puts this twist on a different footing compared to the dipole and Jordanian twists, for example. In fact, it is known that in the undeformed limit the spin-chain Hamiltonian encodes the anomalous dimensions of single-trace gauge-invariant local operators of the CFT. This picture has to change already for the dipole and Jordanian twists, because the usual Cartan is broken, and the spin-chain Hamiltonian must be identified with a different  generator of the conformal group that survives under the deformation. Our considerations on the string-theory side of the duality in section~\ref{sec:string} show that for the Groenewold-Moyal twist the situation is even more drastic: rather than matching the spin-chain Hamiltonian with a  residual symmetry  generator of the conformal group, it seems that we must identify the Hamiltonian with a non-local charge belonging to the tower of integrable charges computable from the monodromy matrix of the string-theory sigma-model.

Our identification was achieved by the explicit construction of a classical solution generalising that of BMN.
Interestingly, it seems that there is some freedom in the identification of this conserved charge dual to the spin-chain Hamiltonian. In fact, depending on how we identify the deformation parameters $\xi$ of the spin-chain and $\eta$ of the sigma-model, to get agreement between the spin-chain Hamiltonian and the string charge  we have to evaluate the (eigenvalue of the) monodromy matrix at a different value of the spectral parameter. It seems natural to check whether this freedom is lifted once we try to match also the charges of different states, or when taking into account higher orders in the large-$J$ expansion of the spin-chain and of the string-sigma model.
The next-to-leading order calculation for the case of a Jordanian twist was carried out  in~\cite{Driezen:2025izd} under a double-scaling limit in powers of $\xi$ and inverse powers of $J$. Looking at~\eqref{eq:en-gr-st}, it seems that in our case the match should be possible without further expanding in $\xi$.

Let us mention that for most Drinfel'd twist deformations of $\mathcal N=4$ super Yang-Mills\footnote{The construction of Drinfel'd twist deformations of $\mathcal N=4$ super Yang-Mills currently covers twists of the Poincar\'e algebra~\cite{Meier:2023lku}, as well as those involving also dilatations~\cite{Borsato:2025jre}.} it is still unclear how to derive a spin-chain description from gauge-theory calculations. In fact, apart from the $\gamma$-deformation that only gives rise to non-commutativity in the R-symmetry, an explicit derivation of the spin-chain picture was initiated only for the dipole deformation of~\cite{Guica:2017mtd} and for the angular dipole deformation of~\cite{Meier:2025tjq}. 
Given that the $CFT_2$ dual to strings on $AdS_3\times S^3\times T^4$ is not yet completely understood, it seems convenient to study the Groenewold-Moyal deformation of $\mathcal N=4$ super Yang-Mills and try to derive a spin-chain description to be compared with the Drinfel'd twist of the undeformed spin-chain. Ideally, this spin-chain should arise, as in the undeformed case, when computing 2-point functions of gauge-invariant operators. It is known how to construct gauge-invariant operators in the presence of the Groenewold-Moyal twist, because following the prescription of~\cite{Gross:2000ba} one should dress traces of star-products of fields with fine-tuned Wilson lines that restore the gauge invariance. A possible future direction is therefore to explicitly calculate 2-point functions of families of these gauge-invariant operators, in the hope of finding an emergent spin-chain description as in the undeformed case. The explicit derivation of the spin-chain would be extremely interesting because, among other things, it would clarify the interpretation of the charge that should be identified with the spin-chain Hamiltonian.

\section*{Acknowledgements}
We would like to thank Sibylle Driezen, Tim Meier, Juan Miguel Nieto Garc\'ia and Stijn van Tongeren for useful discussions, and Sibylle Driezen, Juan Miguel Nieto Garc\'ia and Stijn van Tongeren for comments on the draft.
The work of RB was supported by  RYC2021-032371-I, funded by MCIN/AEI/10.13039/501100011033 and by the European Union ``NextGenerationEU''/PRTR).
The work of MGF was funded by Xunta de Galicia through the ``Programa de
axudas \'a etapa predoutoral da Xunta de Galicia'' (Consellería de Cultura, Educaci\'on
e Universidade) with reference code ED481A-2024-096. 
We also acknowledge the grants 2023-PG083 (with reference code ED431F 2023/19 funded by Xunta de Galicia),  PID2023-152148NB-I00 (funded by AEI-Spain), the Mar\'ia de Maeztu grant CEX2023-001318-M (funded by MICIU/AEI /10.13039/501100011033), the CIGUS Network of Research Centres, and the European Union. 

\appendix

\section{The  $\mathfrak{sl}(2)_L \oplus \mathfrak{sl}(2)_R$ algebra}
\label{ap:sl_2-sl_2}

In this appendix, we review the  $\mathfrak{sl}(2)_L \oplus \mathfrak{sl}(2)_R$ Lie algebra. We adopt the conventions of~\cite{Beisert:2003jj} on the $\mathfrak{sl}(2)$ algebra. Let us denote by the subscripts $L$ and $R$ the generators belonging to each copy of $\mathfrak{sl}(2)$. The algebra is spanned by the basis $\{J^3_L,J^{\pm}_L,J^3_R,J^{\pm}_R\}$, subject to the commutation relations
\begin{align}
    [J^3_M,J^\pm_N] = \pm \delta_{MN} J^{\pm}_M, \quad [J^+_M,J^-_N] = -2 \delta_{MN} J^{3}_M, \quad M,N\in\{L,R\}  \label{eq:algebra_basis}
\end{align}
Moreover, there exist two independent quadratic Casimir operators, one for each copy of $\mathfrak{sl}(2)$, given by
\begin{align}
    C_N=\left(J^3_N\right)^2-\frac{1}{2}\{J^+_N,J^-_N\}, \quad N\in\{L,R\}.
\end{align}

The irreducible representations of  $\mathfrak{sl}(2)_L \oplus \mathfrak{sl}(2)_R$ are labelled by the pair of numbers $(j_L,j_R)$, implicitly defined by the Dynkin labels $[2j_L,2j_R]$ of the representation. We are interested in the infinite-dimensional representation $(j_L,j_R)=(-1/2,-1/2)$. It can be realised by introducing two independent sets of bosonic oscillators $\{a^\dagger,a\}$, $\{\bar a^\dagger,\bar a\}$ such that the only non-zero commutation relations are
\begin{align}
     [a,a^\dagger]=1, \quad [\bar a,\bar a^\dagger]=1.
\end{align}
With this,
\begin{align}\label{eq:osc-rep-sl2}
    J^3_L=\frac{1}{2}+a^\dagger a, \quad J^+_L = a^\dagger+a^\dagger a^\dagger a, \quad J^-_L=a, \\ \nonumber
    J^3_R=\frac{1}{2}+\bar a^\dagger \bar a, \quad J^+_R = \bar a^\dagger+\bar a^\dagger \bar a^\dagger \bar a, \quad J^-_R=\bar a.
\end{align}
The Verma module $V_F$, on which the operators act, is given by the tensor product of two Fock spaces, $V_F^L$ and $V_F^R$, each one corresponding to the $-1/2$ module of the $L$ and $R$ copies of the algebra,
\begin{align}
    V_F^L = \text{span}\{(a^\dagger)^{n}\ket{0}_L \mid n \in \mathbb{N}\}, \quad V_F^R = \text{span}\{(\bar a^\dagger)^{\bar n}\ket{0}_R \mid \bar n \in \mathbb{N}\}, \label{eq:moduleVFN}
\end{align}
\begin{equation}
V_F= V_F^L \otimes V_F^R = \text{span}\{(a^\dagger)^{n}(\bar a^\dagger)^{\bar n}\ket{0} \mid n,\bar n \in \mathbb{N}\}, \label{eq:moduleVF}
\end{equation}
where $a\ket{0}_L=\bar a\ket{0}_R=0$ and we have defined $\ket{0}:=\ket{0}_L \otimes \ket{0}_R$.

Another equivalent realisation of the above representation is the following
\begin{align}
    J_L^3 = z \partial_z+\frac{1}{2}, \quad J^+_L = z^2\partial_z+z, \quad J^-_L = \partial_z, \nonumber \\
    J_R^3 = \bar z \partial_{\bar z}+\frac{1}{2}, \quad J^+_R = \bar z^2\partial_{\bar z}+\bar z, \quad J^-_R = \partial_{\bar z}. \label{eq:hol-rep}
\end{align}
In this realisation, the generators of the algebra must be understood as differential operators acting on the space of analytic functions in the variables $z$ and $\bar z$, corresponding to the $L$ and $R$ copies, respectively. The map between the Fock states and the functions of  $z$ and $\bar z$ is immediate
\begin{align}
    (a^\dagger)^{n}(\bar a^\dagger)^{\bar n}\ket{0} \leftrightarrow z^{n}\bar z^{\bar n}.
\end{align}

Now, we endow  $\mathfrak{sl}(2)_L \oplus \mathfrak{sl}(2)_R$ with a coalgebra structure by defining the coproduct, which allows us to construct the action of the algebra on the tensor product of two modules. In particular, we consider the following primitive coproduct,
\begin{align}
    \Delta(x) = x \otimes\mathbb{I}+\mathbb{I}\otimes x, \quad x \in \{J^3_L,J^{\pm}_L,J^3_R,J^{\pm}_R\}.
\end{align}
Given this choice of the coproduct, the tensor product of two modules $V_F$ decomposes into the following irreducible modules
\begin{align}
    V_F \otimes V_F = \bigoplus_{k,\bar k=0}^{\infty}V_{k,\bar k}, \label{eq:VF-VF} 
\end{align}
where the lowest-weight state of $V_{k,\bar k}$ is 
\begin{align}
    \ket{\Phi_{k,\bar k}}=(a^\dagger_1-a^\dagger_2)^k(\bar a^\dagger_1-\bar a^\dagger_2)^{\bar k} \ket{00}, \quad  \Delta(J_L^-)\ket{\Phi_{k,\bar k}} = \Delta(J_R^-)\ket{\Phi_{k,\bar k}} = 0.
\end{align}
with the subscripts $\{1,2\}$ denoting on which site of the tensor product the operators act. The modules $V_{k,\bar k}$ are eigenspaces of the coproduct of the quadratic Casimir operators,
\begin{align}
    \Delta(C_L) \ket{\Phi_{k,\bar k}} = k(k+1) \ket{\Phi_{k,\bar k}}, \quad \Delta(C_R) \ket{\Phi_{k,\bar k}} = \bar k(\bar k+1) \ket{\Phi_{k,\bar k}}.
\end{align}
From this, one can construct the projectors to the modules $V_{k,\bar k}$ as follows,
\begin{align}
    \mathcal{P}_{k,\bar k} =\mathcal{P}^{L}_k\mathcal{P}^{R}_{\bar k}, \quad \mathcal{P}^{N}_l= \underset{r\neq l}{\prod_{r=0}^\infty} \frac{\Delta(C_N)-r(r+1)}{l(l+1)-r(r+1)}, \label{eq:projectors}
\end{align}
such that $\mathcal{P}_{k,\bar k}V_{r,\bar r}= \delta_{k,r}\delta_{\bar k,\bar r}V_{r,\bar r}$. Moreover, $\mathcal{P}^{L}$ are the projectors acting non-trivially only on the $L$ copy of the algebra, $\mathcal{P}^L_{k}V_{r,\bar r}= \delta_{k,r}V_{r,\bar r}$, and identically for the $R$ copy.

Moreover, let us introduce the $n$-fold coproduct. It is a map $\Delta^{(n)}:V_F \rightarrow V_F^{\otimes n}$ defined recursively as
\begin{align}
    \Delta^{(n)} = \left(\Delta \otimes \mathbb{I}\right) \Delta^{(n-1)}, \quad \text{with} \quad \Delta^{(2)}=\Delta.
\end{align}
This definition allows one to construct the action of the algebra on the $n$-tensor product of $V_F$.

Finally, let us also point out that the $\mathfrak{sl}(2)$ algebra has the automorphism  $J^+\leftrightarrow J^-, J^3\to -J^3$ that we may use for example to write  representations alternative to~\eqref{eq:osc-rep-sl2} and~\eqref{eq:hol-rep}. In fact, in section~\ref{sec:string} we use this automorphism to work with a more convenient $r$-matrix.

\section{The $\mathfrak{sl}(2)_L \oplus \mathfrak{sl}(2)_R$ generators and the Yang--Baxter operators}
\label{ap:comutation-relations}
In this appendix, we derive the commutation relations between the global generators of~$\mathfrak{sl}(2)_L \oplus \mathfrak{sl}(2)_R$ and the Yang-Baxter operators of the $XXX_{-1/2}^{\oplus 2}$ spin-chain. We follow the proof of~\cite{Faddeev:1996iy}.

First, notice that the commutation relations between $L$ and $R$ operators vanish. Let us consider the $RTT$ relation for one copy $N$ of the model
\begin{align}
    R_{a_1, a_2}(u-v) T^N_{a_1}(u) T^N_{a_2}(v) = T^N_{a_2}(v) T^N_{a_1}(u) R_{a_1, a_2}(u-v),\label{eq:RTT}
\end{align}
with auxiliary space $\mathbb{C}^2$. In our conventions, the $\mathfrak{sl}(2)$-invariant $R$-matrix in the $(1/2)$ representation is
\begin{align}
    R_{a_1,a_2}(u) = \left(u+\frac{1}{2}\right)\mathbb{I} + \Pi_{a_1,a_2} \quad \text{with} \quad \Pi_{a_1,a_2}=\frac{1}{2}\sigma^3_{a_1} \otimes \sigma^3_{a_2} + \sigma^+_{a_1} \otimes \sigma^-_{a_2} + \sigma^-_{a_1} \otimes \sigma^+_{a_2}.
\end{align}
Therefore, taking the limit $v \to \infty$ in~\eqref{eq:RTT} leads to
\begin{align}
    &\left(u-v+\frac{1}{2}+\Pi_{a_1,a_2}\right) T^N_{a_1}(u) \left(v^J+v^{J-1}\sum_{n=1}^{J}(\frac{1}{2}\mathbb{I}+\Pi^N_{a_2,n})+O(v^{J-2})\right)=\nonumber \\
    &=\left(v^J+v^{J-1}\sum_{n=1}^{J}(\frac{1}{2}\mathbb{I}+\Pi^N_{a_1,n})+O(v^{J-2})\right) T^N_{a_1}(u) \left(u-v+\frac{1}{2}+\Pi_{a_1,a_2}\right),
\end{align}
where we have defined
\begin{align}
    \Pi_{a,n}^N = \sigma^3_{a} \otimes \left(J^3_{N}\right)_n + \sigma^+_{a} \otimes \left(J^-_{N}\right)_n + \sigma^-_{a} \otimes \left(J^+_{N}\right)_n.
\end{align}
The order $O(v^J)$ cancels, while the order $O(v^{J-1})$ gives the equation
\begin{align}
    \left[T^N_{a_1}(u),\Pi_{a_1,a_2}+\sum_{n=1}^J\Pi^N_{a_2,n}\right] = 0.
\end{align}
Therefore, defining the global $\mathfrak{sl}(2)_N$ generators
\begin{align}
    S^3_N = \Delta^{(J)}(J^3_N), \quad S^{\pm}_N = \Delta^{(J)}(J^{\pm}_N) 
\end{align}
we have
\begin{align}
    \left[S^3_N,T_{a_1}(u)\right]&=\frac{1}{2}\left[T_{a_1}(u),\sigma^{3}_{a_1}\right], \quad \left[S^-_N,T_{a_1}(u)\right]=\left[T_{a_1}(u),\sigma^{-}_{a_1}\right], \nonumber \\
    \left[S^+_N,T_{a_1}(u)\right]&=-\left[T_{a_1}(u),\sigma^{+}_{a_1}\right].
\end{align}
Writing the above equations in a matrix form in $\mathbb{C}^4$ and equating entry by entry, we obtain the following commutation relations
\begin{align}
    \left[S^3_N,A_N\right]&=0,& \left[S^3_N,B_N\right]&=-B_N, &  \left[S^3_N,C_N\right]&=C_N, &  \left[S^3_N,D_N\right]&=0, \nonumber \\
    \left[S^-_N,A_N\right]&=B_N, & \left[S^-_N,B_N\right]&=0, &\left[S^-_N,C_N\right]&=D_N-A_N, &  \left[S^-_N,D_N\right]&=-B_N, \nonumber\\
    \left[S^+_N,A_N\right]&=C_N, & \left[S^+_N,B_N\right]&=D_N-A_N, &  \left[S^+_N,C_N\right]&=0, &  \left[S^+_N,D_N\right]&=-C_N.
\end{align}
Note that, as expected, the transfer  matrix $\tau_N = A_N+D_N$ commutes with all $\mathfrak{sl}(2)_N$ generators.

\section{Computation of $C_N(u)\ket{0}$}
\label{ap:C-action}
In this appendix, we compute the action of the $C_N(u)$ operator on the vacuum for arbitrary values of the spectral parameter $u$. Moreover, we express the result in the basis formed by the one-magnon eigenstates of the undeformed model~\eqref{eq:complete_undef_eigen} and the first descendant of the vacuum.

We start from the $R^N_{a1}$-matrix~\eqref{eq:R-matrix-aux-1/2} with auxiliary space $\mathbb{C}^2$ and acting on the first site of the physical space. Its action on the vacuum is
\begin{align}
    R^N_{a1}(u) \ket{\Omega}= \begin{pmatrix}
        (u+1)\ket{\Omega}& 0 \\
        -\ket{1} & u \ket{\Omega}
    \end{pmatrix},
\end{align}
where $\ket{x}$ with $x \in [1,J]$ denotes a state with one excitation in position $x$. Acting with a second $R$-matrix leads to
\begin{align}
    R^N_{a2}(u) R^N_{a1}(u) \ket{\Omega}= \begin{pmatrix}
        (u+1)^2\ket{\Omega}& 0 \\
        -(u+1)\ket{2}-u\ket{1} & u^2 \ket{\Omega}
    \end{pmatrix},
\end{align}
If we include a third $R$-matrix
\begin{align}
    R^N_{a3}(u) R^N_{a2}(u) R^N_{a1}(u) \ket{\Omega}= \begin{pmatrix}
        (u+1)^3\ket{\Omega}& 0 \\
        -(u+1)^2\ket{3}-u(u+1)\ket{2}-u^2\ket{1} & u^3 \ket{\Omega}
    \end{pmatrix},
\end{align}
from which one can infer the general pattern for the action of $C_N(u)$ on the vacuum
\begin{align}
    C_N(u) \ket{\Omega} = - \frac{u^J}{u+1}\sum_{x=1}^J \left(\frac{u+1}{u}\right)^{x} \ket{x}. \label{eq:C-action}
\end{align}
Notice that if $u$ is a solution of the one-magnon Bethe equation
\begin{align}
    \left(\frac{v_n+1}{v_n}\right)^J =1 \Rightarrow \frac{v_n+1}{v_n} = e^{\frac{2 \pi i n}{J}}, \quad n=1,\ldots,J-1; \label{eq:1-magnon-Bethe-eq}
\end{align}
then
\begin{align}
    C_N(v_n) \ket{\Omega} \propto \sum_{x=1}^J e^{\frac{2 \pi i n}{J}x} \ket{x} \label{eq:C-bethe}
\end{align}
and we recover the plane-wave solution of the coordinate Bethe Ansatz.

Now, we want to express~\eqref{eq:C-action} in the basis
\begin{align}
    \{S_N^+\ket{\Omega}, C_N(v_n)\ket{\Omega}\}, \quad n=1, \ldots J-1.
\end{align}
That is to say, we need to rewrite
\begin{align}
    \ket{x} = \alpha_0(x) S_N^+ \ket{\Omega} + \sum_{n=1}^{J-1} \alpha_n(x) C_N(v_n) \ket{\Omega}, 
\end{align}
for some unknown coefficients $\alpha's$  that in general could depend on the position $x$ under consideration. Using~\eqref{eq:C-bethe}, and defining the normalisation factor of~\eqref{eq:C-action}
\begin{align}
    g(u) = - \frac{u^J}{u+1},
\end{align}
it follows that the $\alpha's$ must be solutions of the following system of equations
\begin{align}
    \alpha_0(x) + \sum_{n=1}^{J-1} \alpha_n(x) g(v_n) e^{\frac{2 \pi i n x}{J}}&=1,\\
    \alpha_0(x) + \sum_{n=1}^{J-1} \alpha_n(x) g(v_n) e^{\frac{2 \pi i n y}{J}}&=0, \quad \text{for} \quad y=1, \ldots,J, \quad \text{and} \quad y \neq x .
\end{align}
To solve the previous system of equations, we perform the following change of variables
\begin{align}
    \tilde{\alpha}_n(x) = \alpha_n(x) g(v_n) e^{\frac{2 \pi i n x}{J}},\quad n=1, \ldots,J-1.
\end{align}
Then, the system can be recast as follows
\begin{align}
    \alpha_0(x) + \sum_{n=1}^{J-1} \tilde{\alpha}_n(x)&=1,\label{eq:alpha_1}\\
    \alpha_0(x) + \sum_{n=1}^{J-1} \tilde{\alpha}_n(x)  e^{\frac{2 \pi i n (x-y)}{J}}&=0, \quad \text{for} \quad y=1, \ldots,J, \quad \text{and} \quad y \neq x. \label{eq:alpha_2}
\end{align}
Now, notice that
\begin{align}
    1 + \sum_{n=1}^{J-1} e^{\frac{2 \pi i n (x-y)}{J}} = \delta_{x,y} J.
\end{align}
Therefore, the equations~\eqref{eq:alpha_2} are solved by $\alpha_0(x) = \tilde{\alpha}_n(x)$, while equation~\eqref{eq:alpha_1} fixes $\alpha_0(x) = \tilde{\alpha}_n(x)=1/J$. Thus, the solution of the system is given by
\begin{align}
    \alpha_0(x) &= \frac{1}{J},\\
    \alpha_n(x) &= \frac{1}{J g(v_n)} e^{-\frac{2 \pi i n x }{J}}, \quad n=1, \ldots , J-1.
\end{align}
With this,~\eqref{eq:C-action} can be written as
\begin{align}
    C_N(u) \ket{\Omega}= \frac{g(u)}{J} \sum_{x=1}^J\left(\frac{u+1}{u}\right)^x\left(S^+_N \ket{\Omega}+\sum_{n=1}^{J-1} \frac{e^{-\frac{2 \pi i n x }{J}}}{g(v_n)} C_N(v_n) \ket{\Omega}\right).
\end{align}
Using the identity
\begin{align}
    \sum_{x=1}^J\left(\frac{u+1}{u}\right)^x = (1+u)\left(\left(\frac{u+1}{u}\right)^J-1\right),
\end{align}
we can simplify the above expression as follows
\begin{align}
    &C_N(u) \ket{\Omega}= \frac{u^J-(u+1)^J}{J} S^+_N \ket{\Omega}+\nonumber \\
    &+\frac{g(u)}{J} \sum_{n=1}^{J-1}\sum_{x=1}^J\left(\frac{u+1}{u}\right)^x \frac{e^{-\frac{2 \pi i n x }{J}}}{g(v_n)} C_N(v_n) \ket{\Omega}.
\end{align}
As a consistency check, we can verify that if we set $u$ to be a solution of the one-magnon Bethe equation~\eqref{eq:1-magnon-Bethe-eq}, $v_k$, we recover $C_N(v_k) \ket{\Omega}$. In fact, the coefficient that multiplies $S^+ \ket{0}$ cancels, while
\begin{align}
    \sum_{x=1}^J \left(\frac{v_k+1}{v_k}\right)^x e^{-\frac{2 \pi i n x }{J}} =  \sum_{x=1}^J e^{\frac{2 \pi i (k-n) x }{J}} = J \delta_{n,k},
\end{align}
which implies that
\begin{align}
    C_N(u \to v_k)\ket{\Omega} = g(v_k)\sum_{n=1}^{J-1} \frac{\delta_{n,k}}{g(v_n)}C_N(v_n) \ket{\Omega} = C_N(v_k) \ket{\Omega}.
\end{align}
For $u\neq v_n$, the coefficient that multiplies $C_N(v_n)$ takes the simple form
\begin{align}
    \frac{g(u)}{J} \sum_{x=1}^{J} \left(\frac{u+1}{u}\right)^x \frac{e^{-\frac{2 \pi i n x }{J}}}{g(v_n)} = \frac{v_n\left(u^J-(u+1)^J\right)}{J (1+v_n)^{J-1} (u-v_n)}.
\end{align}
A special point corresponds to $u \to 0$. In this case, one has
\begin{align}
    C_N(0) \ket{\Omega} = - \ket{J} = - \frac{1}{J} \left(S^+_N \ket{\Omega}-\sum_{n=1}^{J-1}\frac{1}{(1+v_n)^{J-1}} C_N(v_n) \ket{\Omega}\right).
\end{align}
\section{On the transfer matrix eigenvalues at arbitrary values of $u$.}
\label{ap:proo-dif-eigenv}
In this appendix, we prove that eigenstates of the $XXX_{-1/2}$ transfer matrix, evaluated at generic values of $u$, with different magnon numbers and finite Bethe roots have different eigenvalues. We prove this claim by adapting the theorem 3.3 of~\cite{Belliard:2018pvg} to our non-compact representation.

The starting point is the $M$-magnon $Q$ function, defined as
\begin{align}
    Q(u)=\prod_{i=1}^M \frac{(u-v_i)}{c}, \quad M \in \mathbb{N},
\end{align}
where $c$ is some arbitrary constant~\footnote{Notice that one can always set $c=1$ by rescaling the spectral parameter and the Bethe roots as $u \to cu$ and $v_i \to c v_i$. However, for the sake of the proof we keep $c$ arbitrary. } and $v_i$ denotes the Bethe roots defining the Bethe state. The $Q$ function satisfies the so-called $T-Q$ relation, which in our conventions reads
\begin{align}
    \Lambda(u) Q(u) = (u+c)^J Q(u+c) + u^J Q(u-c),
\end{align}
where $\Lambda(u)$ is the transfer matrix eigenvalue of the Bethe state associated with $Q(u)$.

Now suppose there exist two Bethe states with different magnon numbers and Bethe roots but with the same eigenvalue of the transfer matrix. This implies that there must exist two $Q$ functions
\begin{align}
    Q_1(u)=\prod_{i=1}^{M_1} \frac{(u-v_i)}{c}, \quad Q_2(u)= \prod_{i=1}^{M_2} \frac{(u-w_i)}{c},
\end{align}
such that
\begin{align}
    \Lambda(u) Q_1(u) = (u+c)^J Q_1(u+c) + u^J Q_1(u-c), \nonumber \\
    \Lambda(u) Q_2(u) = (u+c)^J Q_2(u+c) + u^J Q_2(u-c), \label{eq:T-Q-1-2}
\end{align}
for the same function $\Lambda(u)$. 
Let us define the quantum Wronskian
\begin{align}
    W(u)=u^J\left[Q_1(u-c) Q_2(u)-Q_1(u) Q_2(u-c)\right]. \label{eq:q-Wronskian}
\end{align}
Using equations~\eqref{eq:T-Q-1-2}, one finds the following identity for the Wronskian,
\begin{align}
    W(u)-W(u+c)=0,
\end{align}
which implies that $W(u)$ is a constant. With this, expanding equation~\eqref{eq:q-Wronskian} in powers of $u$, we find that
\begin{align}
    \frac{W}{u^J} =  \left(M_2-M_1\right) \left(\frac{u}{c}\right)^{M_1+M_2-1} + O\left(u^{M_1+M_2-2}\right),
\end{align}
which implies that $W=0$ and all coefficients on the right-hand side vanish. In particular cancelling the term at order $O\left(u^{M_1+M_2-2}\right)$ gives the condition $M_1=M_2$. That is to say, all Bethe states with the same eigenvalue of the transfer matrix must have the same number of magnons.

Moreover, one can show that they have the same set of Bethe roots. To this end, we evaluate equation~\eqref{eq:q-Wronskian} at a Bethe root $v_i$ of $Q_1(u)$. This yields,
\begin{align}
    Q_1(v_i-c) Q_2(v_i)=0 \Rightarrow Q_2(v_i)=0.    
\end{align}
Therefore, all Bethe roots of $Q_1$ are roots of $Q_2$. Repeating the same argument with the Bethe roots of $Q_2$, we conclude that the converse is also true, which proves the claim.
\bibliographystyle{nb}
\bibliography{biblio}{}
\end{document}